\RequirePackage{lineno} 

\documentclass[12pt,preprint]{aastex}

\usepackage{natbib}
\usepackage{color}
\usepackage{multirow}

\shorttitle{3D GCM Intercomparison}
\shortauthors{Yang et al.}

\begin{document}



\title{Ocean Dynamics and the Inner Edge of the Habitable Zone for Tidally Locked Terrestrial Planets}

\author{Jun Yang$^{1}$, Dorian S. Abbot$^2$, Daniel D. B. Koll$^3$, 
Yongyun Hu$^1$, and Adam P. Showman$^{4,1}$}
  \affil{$^1$Dept. of Atmospheric and Oceanic
  Sciences, School of Physics, Peking University, Beijing, 100871, China
  \\  $^2$Dept. of the Geophysical Sciences, University of Chicago, Chicago, IL, 60637, USA\\ 
  $^3$Dept. of Earth, Atmospheric and Planetary Sciences, MIT, Cambridge, MA, 02139, USA\\ 
  $^4$Dept. of Planetary Sciences and Lunar and Planetary Laboratory, University of Arizona, AZ, 85721, USA\\}
\email{Correspondence:~junyang@pku.edu.cn}

\begin{abstract}
Recent studies have shown that ocean dynamics can have a significant warming effect 
on the permanent night sides of 1:1 tidally locked terrestrial exoplanets with Earth-like atmospheres 
and oceans in the middle of the habitable zone. However, the impact of ocean dynamics on 
the habitable zone’s boundaries (inner edge and outer edge) is still unknown 
and represents a major gap in our understanding of this type of planets. Here we 
use a coupled atmosphere-ocean global climate model to show that planetary heat 
transport from the day to night side is dominated by the ocean at lower stellar fluxes 
and by the atmosphere near the inner edge of the habitable zone. This decrease 
in oceanic heat transport (OHT) at high stellar fluxes is 
mainly due to weakening of 
surface wind stress and a decrease in surface shortwave energy deposition. We further 
show that ocean dynamics have almost no effect on the observational thermal 
phase curves of planets near the inner edge of the habitable zone. For planets in the 
habitable zone’s middle range, ocean dynamics moves the hottest spot on the surface 
eastward from the substellar point. These results suggest that future studies of the 
inner edge may devote computational resources to atmosphere-only processes 
such as clouds and radiation. For studies of the middle range and outer edge of the habitable zone, 
however, fully coupled ocean-atmosphere modeling will be necessary. Note that due 
to computational resource limitations, only one rotation period (60 Earth days) has 
been systematically examined in this study; future work varying rotation 
period as well as 
other parameters such as atmospheric mass and composition is required. 
\end{abstract}

\keywords{astrobiology  --- planets and satellites: oceans --- planets and satellites: terrestrial planets 
--- methods: numerical --- stars: low-mass }


\section{Introduction}
\label{sec:introduction}

The ocean has a profound effect on the variation and time-mean features of the climate of Earth 
through modifying the surface heat capacity, transporting heat from low latitudes to high latitudes 
and storing carbon \citep{vallis2012climate,Watsonetal2015}. The tight interaction between 
ocean, atmosphere, ice and clouds further influence global and regional energy balances and 
surface temperatures. For instance, if Earth’s global ocean circulation were artificially turned off, 
the global-mean surface temperature would decrease by several degrees \citep{Winton2003JCli,
HERWEIJER:2005bq}. How ocean dynamics influence the climate and habitability of 
exoplanets remains relatively unstudied.

During the past 20 years, various three-dimensional (3D) atmosphere-only climate models have 
been employed to examine the important effects of atmospheric circulation on the climate and 
habitability of 1:1 tidally locked (`synchronously rotating') terrestrial planets 
\citep[e.g.,][]{Joshi:1997,Merlis:2010,Edson:2011p2367,Pierrehumbert:2011p3287,
leconte2013increased,Leconte:2013gv,yang2013,yang2014b,Wangetal2014:eccentric,
Wangetal2016:obliquity,way2015exploring,kopparapu2016inner,popp2016,
turbet2016habitability,salameh2017role,wolf2017assessing,wolf2017constraints,
haqq2017demarcating,boutle2017exploring,kopparapu2017habitable,binetal2018,turbet2018modeling}. 
These studies employed a dry land surface with no ocean or an immobile thermodynamic 
ocean with no ocean dynamics. The effect of ocean dynamics on exoplanets has only 
been addressed by a few studies \citep{yang2013,yang2014b,hu2014role,Cullumetal2014rotation,Cullumetal2016salinity,
way2015exploring,Wayetal2018Ocean,Delgenioetal2017:Proxima_b}. Studies on synchronously rotating planets 
have found that ocean dynamics could significantly 
warm the permanent night sides of planets in the middle range of the habitable zone 
(\cite{hu2014role} and \cite{Delgenioetal2017:Proxima_b}, or see section \ref{OHTMiddleRange} 
for more detailed feedback analyses). A critical unaddressed question is: Could ocean dynamics 
have a significant effect on the location of the inner edge of the habitable zone? If ocean dynamics 
have a strong warming effect on planets near the inner edge, as they do for those in the 
middle range, this could shrink the habitable zone. As a result, the number of potentially habitable 
planets would be lower, making it harder to detect and study such planets in the future. 
Here we try to answer this question through a series of 3D climate experiments. 


The maximum stellar flux 
investigated by \cite{hu2014role}, 1,400 W m$^{-2}$, is only about 40-50\,\% of the stellar flux 
at the inner edge, according to results from atmosphere-only models \citep{yang2014b,kopparapu2017habitable,binetal2018}. 
Our new experiments show that the day-to-night oceanic heat transport 
(OHT) at the higher stellar fluxes of planets near the inner edge is much weaker than that for 
a stellar flux of 1,400~W m$^{-2}$ (see section~\ref{OHTtrend}) and that the location of the inner 
edge would not be shifted by ocean dynamics, at least for a rotation period 
of 60~Earth days. Section~\ref{sec:ModelandExperiments} addresses the model description and 
experimental design. We show the results in section~3, including the effect of ocean dynamics on 
planets in the middle range (section~3.1) and near the inner edge (section~3.2) of the habitable 
zone, the effect of ocean dynamics on observational thermal phase curves (section~3.3) and 
an exception to our result that OHT tends to monotonically decrease with 
increasing stellar flux (section~3.4). We further discuss the results and suggest future required 
studies in section~4, and summarize in section~5.



\section{Model Description and Experimental Design}
\label{sec:ModelandExperiments}

We use the Community Climate System Model CCSM3, which has four coupled 
components: atmosphere, ocean, land and sea ice \citep{collins2006ccsm3}. 
The model was developed to investigate the climates of present, past and future Earth. 
We have modified the model to be able to simulate the climates of Earth-like exoplanets 
that have different stellar spectra, planetary orbits, masses, and land-sea distributions 
\citep{Liu:2013gm,hu2014role,yang2014b}. By default, we use a planetary surface 
nearly completely covered with an ocean of water (“an aqua-planet”) with a uniform 
depth of $\simeq$\,4,000 m, close to the mean depth of Earth’s oceans. We include two small 
islands at the south and north poles (poleward of 85\,$^{\circ}$S(N)) because the poles 
of the ocean grid have to reside on continents \citep{RosenbloometalCCSM3}. 
These small-area islands should have a negligible influence on the 
results\footnote{\cite{Delgenioetal2017:Proxima_b} used another coupled 
atmosphere-ocean circulation model ROCKE-3D to examine the possible climate 
scenarios of a tidally locked planet---Proxima Centauri b 
and employed an aqua-planet configuration without any continent in most of their simulations. 
The main characteristics of ocean circulation (ocean currents, spatial patterns of sea 
surface temperature and sea ice, etc.) in their simulations are similar to our results.}. 
Aqua-planet simulations provide a standard framework for understanding large-scale 
ocean circulations and for relating our results to previous work, including analytical theories  \citep{Smithetal2006:aquaplanet,Marshalletal2007:aquaplanet}. 
Also, due to the lack of complex land barriers, an aqua-planet represents the 
simplest fully 3D and coupled system in which to investigate ocean circulation and 
ocean-atmosphere interaction, and therefore is most appropriate for examining the 
mechanisms that strengthen or weaken the ocean's effect at different stellar fluxes.



The atmosphere is assumed to be Earth-like, composed of N$_2$ and H$_2$O, with a surface 
pressure of approximately 1.0 bar (depending on the water vapor concentration). The 
concentrations of CO$_2$, CH$_4$, N$_2$O, O$_2$, O$_3$ and CFCs are 
set to zero. Water vapor is the only greenhouse gas. We chose one sample 
star with a surface temperature of 4,500 K. The rotation period of the planet 
is 60 Earth days ($=$ orbital period), and the stellar fluxes we tested are 1,400, 
1,600, 1,800, 2,000, 2,200, 2,400, 2,600, and 2,800 W m$^{-2}$.
 The maximum stellar flux for which the model can achieve a quasi-equilibrium state 
 is 2,800 W m$^{-2}$. The radius and gravity of the simulated planet are set to 
 typical values of a super-Earth (such as the unconfirmed exoplanet Gl 581g 
 \citep{Vogt:2010p3247} and the confirmed exoplanet LHS 1140b \citep{Dittmannetal2017:LHS1140b}):  
 1.5 times Earth’s radius and 1.38 times Earth’s gravity. Both obliquity and 
 eccentricity are set to zero. 
 
 
\begin{table}[!htbp]
\label{Table:ExperimentalDesign}
\begin{center}
  \caption{Summary of the climate simulations performed using CCSM3}
\begin{tabular}{p{2.2cm}cp{12cm}}
  \tableline\tableline\noalign{\smallskip}
  Group & Runs & Experimental Design  \\
  \tableline\noalign{\smallskip}
Control  & 1 & The surface is covered by an ocean with a uniform depth of about 4,000 m. 
The stellar flux is 1,400 W m$^{-2}$ with a 4,500-K blackbody spectrum. 
Rotation period ($=$ orbital period) is 60 Earth days. \\
\hline
Stellar flux  & 9 & Same as `Control' except the stellar flux is varied between 
1,400 and 2,800 W m$^{-2}$ in increments of 200 W m$^{-2}$. \\
\hline
Initial state  & 2 & Same as `Control' except stellar flux is set to 1,800 W m$^{-2}$ and 
the initial state is from the equilibrium state of the case of 2,400 W m$^{-2}$, or the stellar 
flux is set to 2,400 W m$^{-2}$ and the initial state is from the equilibrium state of the 
case of 1,800 W m$^{-2}$. \\
\hline
Ocean depth & 3 & The ocean depth is set to 4,000, 800 and 400 m, respectively. 
The stellar flux is 1,400 W m$^{-2}$ with a 3,400-K blackbody spectrum. 
Rotation period ($=$ orbital period) is 37 Earth days. \\
\hline
Land-sea distribution   & 2 & One-ridge world: a thin barrier that completely obstructs ocean 
flows running from pole to pole on the eastern terminator. Two-ridges world: two thin barriers 
on the western and eastern terminators. Other parameters are the same as the ‘Ocean depth’ cases.\\
  \tableline\noalign{\smallskip}
  \end{tabular}
  \end{center} 
\end{table}

In our simulations, for a given stellar temperature, the rotation period is fixed when 
varying the stellar flux, as was done by \cite{Wangetal2014:eccentric}, \cite{way2015exploring}, 
and \cite{Fujiietal2017:Moist}. The period of 60 Earth days we chose is 
close to the rotation period near the inner edge of the habitable zone for a star 
with a temperature of 4,500~K (see Fig.~2 of ref. \cite{kopparapu2016inner}). 
This design allows us to isolate the effect of increasing stellar radiation 
and is sufficient to demonstrate the role of ocean dynamics on the inner edge 
of the habitable zone. Our setup is different from that employed by 
\cite{kopparapu2016inner}, \cite{haqq2017demarcating} and \cite{wolf2017constraints}, 
who modified the rotation 
period and stellar flux simultaneously. Their experiments are able to self-consistently 
consider the combined effect of the Coriolis force and stellar flux, but do not allow 
the separate consideration of each factor. In the future, it would be interesting to 
investigate the combined effect of varying the rotation period and stellar flux using 
a coupled atmosphere--ocean model.

 In order to test the effect of continent and ocean depth on the ocean circulation, we also carried out 
 experiments with a one-ridge or a two-ridges continental setup\footnote{The ocean depth and the land-sea distribution experiments were run about 3 years ago, while the other experiments were run this year and employed somewhat different stellar temperatures and rotation periods (Table~1). However, the differences in stellar temperature and rotation period should have a very small effect on the results here and don't influence the conclusion of this paper. This is due to that fact that (1) both 37 days and 60 days are in the slowly rotating regime \citep{Edson:2011p2367,haqq2017demarcating} and (2) although the stellar spectrum can influence the surface albedo of ice and snow, the day side is ice-free in most of the CCSM3's experiments.} or different 
 ocean depths, 800 and 400 m (Table \ref{Table:ExperimentalDesign}). The geothermal heat 
 flux is set to zero, and the only energy source is stellar radiation. We find our results 
 are robust across these parameters but future work is required to consider whether 
 our results are strongly sensitive to additional effects, such as other 
 resonant orbital-rotational states, different planetary 
 radii and gravities, different land-sea distributions, and other ocean depths.

\begin{figure}[!htbp]
\begin{center}
\includegraphics[angle=0, width=16cm]{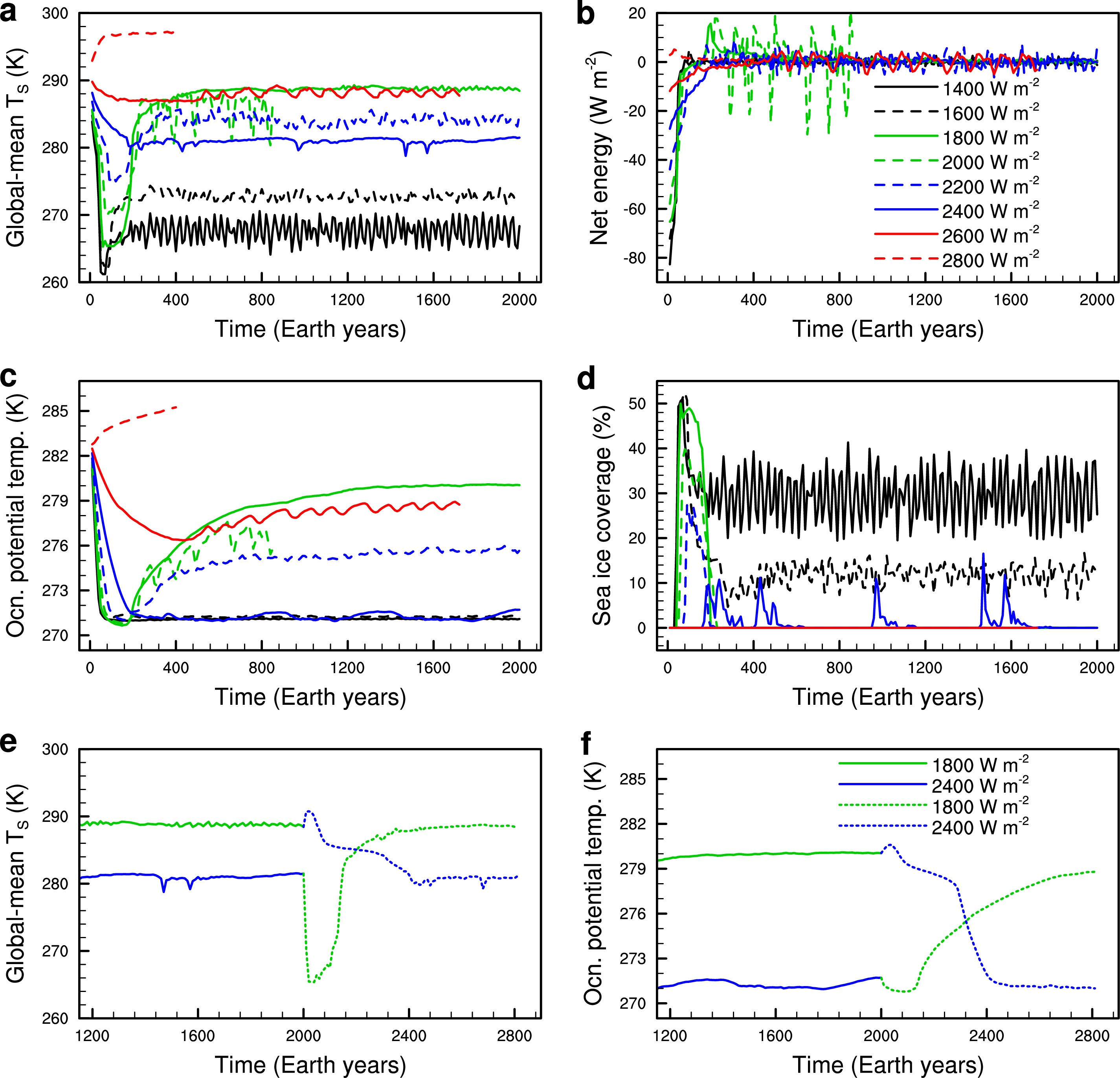}
\caption{Time series of global-mean characteristics of the CCSM3 simulations. 
(a), surface air temperature; (b), net energy flux (shortwave minus longwave) at the top 
of the atmosphere; (c), vertically-averaged ocean potential temperature; and (d), sea ice coverage. 
The stellar flux is increased from 1,400 to 2,800 W m$^{-2}$ in increments of 200 W m$^{-2}$. 
(e-f), independence of the equilibrium state from the initial state, for the stellar fluxes of 1,800 
and 2,400 W m$^{-2}$. (e), global-mean surface temperature; and (f), global- and vertical-mean 
ocean potential temperature. In (e-f), the solid lines are the same as those shown in (a) and (c), 
and the dotted lines are initialized from the equilibrium states of the solid lines. Each data point 
in these time series is the average over 10 Earth years. In all of these experiments, the rotation 
period is 60 Earth days and we apply a 4,500-K blackbody spectrum and an aqua-planet with 
a uniform depth of $\simeq$\,4000 m.}
\label{Figure:TimeSeries}
\end{center}
\end{figure}

In the ocean component of our model, diffusion and viscosity parameters are assumed to be 
the same as for Earth’s present-day oceans. The parameterization of along-isopycnal 
potential temperature and salinity diffusion uses the Gent-McWilliams scheme 
\citep{Gent-Mcwilliams-1990:isopycnal} with a coefficient of 800 m$^2$ s$^{-1}$. 
Horizontal viscosity in the momentum equation employs the anisotropic 
formulation \citep{SmithandMcWilliams2003}. Diapycnal mixing is represented 
by the K-profile parameterization boundary-layer scheme 
\citep{Large-Mcwilliams-Doney-1990:oceanic}. Several physical processes are 
considered in the scheme, including internal waves, shear instability, convective 
instability and double diffusion \citep{SmithandGent2004}. Separate studies 
using high-resolution eddy-resolving ocean models are required to understand 
the uncertainties of these parameters for Earth-like exoplanets; computational resource 
limitations preclude such sensitivity experiments at present.

The atmosphere and land components of the model have a horizontal resolution 
of 3.75$^{\circ}$ $\times$ 3.75$^{\circ}$ and with 26 vertical levels from the 
surface to $\sim$36 km. The ocean and sea-ice components have a variable 
latitudinal resolution starting at 0.9$^{\circ}$ near the equator, a constant 
longitudinal resolution of 3.6$^{\circ}$, and 25 vertical levels. By default, the 
time step for the atmosphere and land components is set to 900 s. For the 
simulations with high stellar fluxes, we used a smaller time step to avoid 
numerical instability; the minimum time step we examined is 60 s. Due to 
computational limitations, time steps less than 60 s were not tested. 
For the ocean, the time step is two hours. The coupling time interval between 
the atmosphere and ocean is one Earth day. The sea ice albedo is 0.50 in the 
visible and 0.30 in the near infrared. The snow 
albedo is 0.91 in the visible and 0.63 in the near infrared 
\citep{Briegleb-et-al-2002:description}. The broadband surface albedo, therefore, 
depends on stellar spectrum and is less reflective at redder 
wavelengths \citep{Joshi:2012hu,shields2013effect}.

The atmosphere was initialized from a state close to modern Earth, and the ocean 
was initialized from a state of rest with a horizontally uniform temperature. We 
integrated each case for about 1,000 or 2,000 Earth years, and present the final 100 or 
200 years in the following analyses. Time series of global-mean surface temperature, 
energy balance (absorbed shortwave radiation minus outgoing longwave radiation) 
at the top of the atmosphere, vertically averaged ocean potential temperature, and 
sea ice coverage are shown in Fig.~\ref{Figure:TimeSeries}(a-d). In most cases, the 
system reaches quasi-equilibrium within about 1,000 years, although several simulations exhibit  
significant oscillations after that time: (1) The 2,000 and 2,600 W\,m$^{-2}$ cases show 
strong oscillations with a period of $\simeq$\,100 Earth years; (2) the 1,400 and 
1,600 W\,m$^{-2}$ cases also exhibit significant variations but with a period of 
$\simeq$\,30 Earth years; (3) the 2,400 W\,m$^{-2}$ case shows irregular risings 
in sea ice coverage and fallings in sea temperature and ocean potential temperature. 
We speculate that these long-time oscillations may arise from the interactions 
between atmosphere, sea ice and ocean (especially Rossby and Kelvin waves), 
such as that shown in \cite{Marshalletal2007:aquaplanet}; 
detailed analyses of the underlying mechanisms are beyond the 
scope of the present work. In this paper, we will focus on the mean state only.
Note that ocean potential temperature in the 2,800 W\,m$^{-2}$ case is still increasing 
although the surface temperature does not show a significant warming trend; the model 
blew up when we tried to further integrate the run. If the model could be integrated longer, 
the ocean would become warmer especially in relatively cold regions at the ocean bottom 
and the ocean in the night side, so that OHT from the day side to the night side (see 
section~\ref{OHTtrend} below) would likely be even smaller. 




In the following section, we will analyze our equilibrated CCSM3 simulations to understand 
the ocean's effect on the planetary climates. To isolate the effect of ocean dynamics, we preformed 
corresponding atmosphere-only experiments using the Community Atmosphere Model 
version 3.1 (CAM3, \cite{collins2004description}), which is the atmosphere component 
of CCSM3. CAM3 is coupled with a 50-m mixed layer, immobile ocean, and ocean heat 
transport is specified to be zero everywhere. The model is coupled to a thermodynamic 
sea ice model, in which sea ice flows are not considered. Each case is integrated for 
about 60 Earth years and reaches a steady state after about 40 years. Averages over the 
last 5 years are used for our analyses. The maximum stellar flux for which simulations in 
CAM3 can achieve a quasi-equilibrium state is 3,000 W\,m$^{-2}$, higher than that in CCSM3.


\section{Results}
\label{sec:results}

\subsection{The Ocean's strong effect in the middle of the habitable zone}
\label{OHTMiddleRange}

\begin{figure}[!htbp]
\begin{center}
\includegraphics[angle=0, width=16cm]{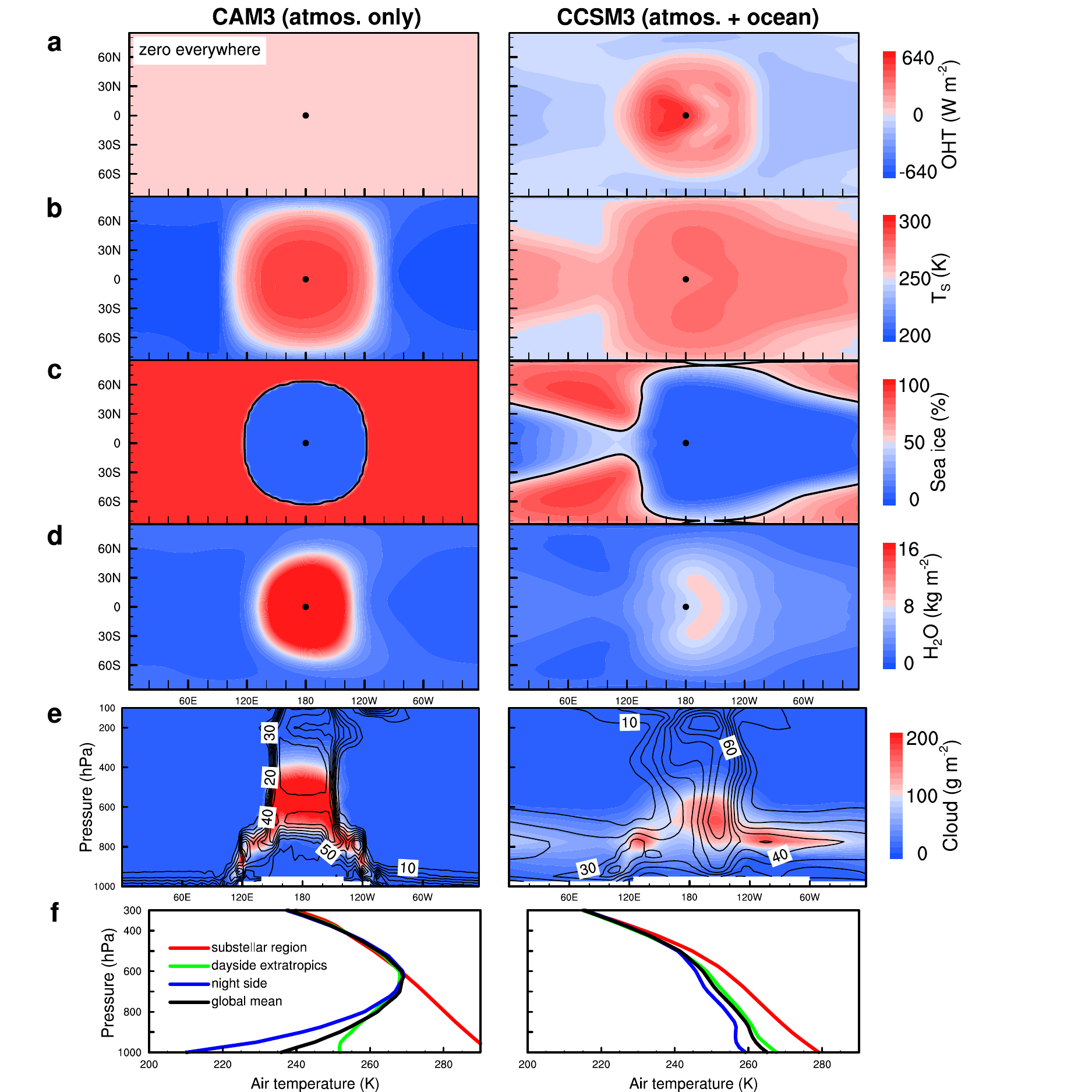}
\caption{Strong effects of ocean dynamics on a tidally locked aqua-planet in the middle range 
of the habitable zone. Left: CAM3, and right: CCSM3. (a), OHT (i.e., surface net heat flux; 
negative: heat from ocean to atmosphere, positive: heat from atmosphere to ocean); 
(b), surface temperature; (c), sea ice coverage (the black line shows 50\,\% coverage); 
(d), vertically integrated water vapor amount; (e), meridional-mean cloud water amount (color 
shading) and cloud fraction (\%, contour lines with an interval of 10\,\%); and (f), 
vertical temperature profiles for different regions. 
Note the vertical axis is latitude in (a-d) but air pressure in (e-f). In both experiments, the stellar 
flux is 1,400~W~m$^{-2}$. In this figure and elsewhere, the substellar point is at 
(180$^{\circ}$,~0$^{\circ}$) and is marked with a black dot.}
\label{Figure:MiddleRange}
\end{center}
\end{figure}

The ocean transports heat from the substellar region where there is net energy 
gain to the night side where there is net energy loss. As a result, the night side in 
CCSM3 is much warmer than that in CAM3 by 40-50 K (Fig.~\ref{Figure:MiddleRange}; 
see also \cite{hu2014role} and \cite{Delgenioetal2017:Proxima_b}). The ocean also 
transports heat in the north-south direction and thereby warms the dayside high 
latitudes. Despite the ocean transporting heat away from the substellar region, 
surface temperatures there decrease only slightly, by $\simeq$\,10 K. This is mainly 
due to a negative cloud feedback: As the OHT carries energy away from the 
substellar region, the surface temperature decreases and therefore convection 
over the substellar region becomes weaker, so that cloud optical depth 
(Fig.~\ref{Figure:MiddleRange}(e)) and planetary albedo decrease, allowing 
more stellar energy to reach the surface and warm it. This feedback is similar to 
that described in \cite{koll2013tropical}. Moreover, per unit of energy, 
the warming of the nightside surface should be greater than the cooling  
of the dayside surface because of the nonlinear dependence 
of thermal emission on temperature (i.e., the Planck effect) and the fact 
that the nightside surface is colder than the dayside surface.

\begin{figure}[!htbp]
\begin{center}
\includegraphics[angle=0, width=16cm]{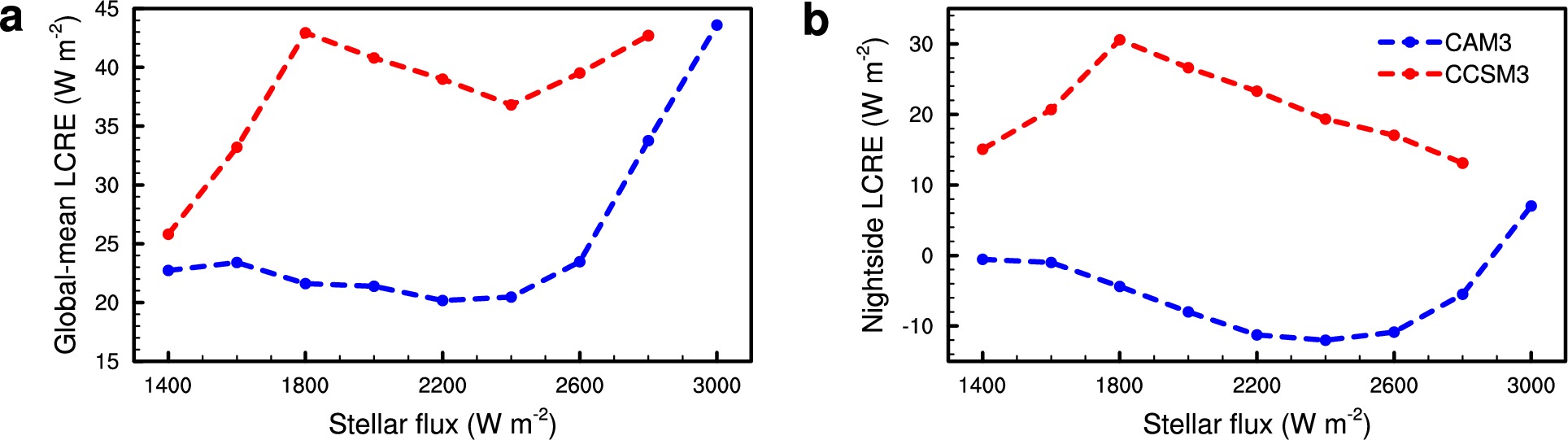}
\caption{Longwave cloud radiative effect (LCRE) at the top of the atmosphere. 
Blue: CAM3, and red: CCSM3. (a), global mean, and (b), nightside average.}
\label{Figure:CloudRE}
\end{center}
\end{figure}

Besides the direct effect of ocean dynamics, feedbacks associated with sea ice, 
water vapor and atmospheric lapse rate further act to warm the surface. OHT 
melts the sea ice around the terminators and at the dayside high 
latitudes (Fig.~\ref{Figure:MiddleRange}(c)), reducing the planetary albedo. 
Meanwhile, in the experiments with stellar fluxes equal to or less than 1,600 W\,m$^{-2}$, 
sea ice in CCSM3 is only several meters thick, being 2-3 orders 
of magnitude thinner than the nightside sea ice thickness in CAM3, similar to the 
results in \cite{yang2014b}, so OHT is effective at preventing water trapping 
on the night side. For stellar fluxes higher than 
1,600 W\,m$^{-2}$, there is no snow or ice either on the day or night side 
in CCSM3, although in CAM3 the night side is still covered by ice and snow. 
Compared to CAM3, CCSM3’s water vapor concentration is higher on the 
night side (Fig.~\ref{Figure:MiddleRange}(d), although lower in the substellar 
region), the temperature inversion (i.e., air temperature being higher than the surface, 
ref.~\cite{zalucha_investigation_2013,Leconte:2013gv,YangandAbbot2014}) disappears 
on the night side (Fig.~\ref{Figure:MiddleRange}(f)), and the longwave 
cloud radiative effect is higher (a warming effect, Fig.~\ref{Figure:CloudRE}). 
All of these factors act to increase the atmospheric greenhouse effect on the night side in CCSM3. 
In contrast, on the night side in CAM3, clouds form at the layers near the temperature inversion   
(Fig.~\ref{Figure:MiddleRange}(f)), which acts to trap water vapor that evaporated 
from the surface; this mechanism is similar to the low-level stratus cloud formation 
over the eastern subtropical Pacific Ocean where the sea surface 
is cooler than average sea surface due to ocean upwelling there (ref.~Chapter 3.13 
of \cite{Hartmann-2016:book}). These clouds emit longwave radiation to space 
at temperatures higher than the surface, causing a negative longwave cloud radiative 
effect on the night side (Fig.~\ref{Figure:CloudRE}(b)) and inducing a cooling effect 
on the air and the surface, except the case of 3,000 W\,m$^{-2}$ in which the temperature 
inversion disappears. In CCSM3, there is no temperature inversion in all of the 
experiments shown here (note that for a much lower stellar flux, such as 625 W\,m$^{-2}$ 
in which ocean heat transport is very weak, there is a temperature inversion (figure not shown))  
and the longwave cloud radiative effect is positive (warming) everywhere.


The direct effect of ocean dynamics is therefore to reduce the day-to-night surface temperature 
contrast in CCSM3 compared to CAM3. Associated surface and atmospheric feedbacks then 
amplify the nightside warming, which is why global-mean surface temperature in CCSM3 is 
much higher than that in CAM3 (Fig.~\ref{Figure:MiddleRange}(b)). However, as we show in the 
next section, these differences become smaller and smaller as the stellar flux is increased.

\subsection{Decreasing Trend of OHT With Increasing Stellar Flux}
\label{OHTtrend}

\begin{figure}[!htbp]
\begin{center}
\includegraphics[angle=0, width=16cm]{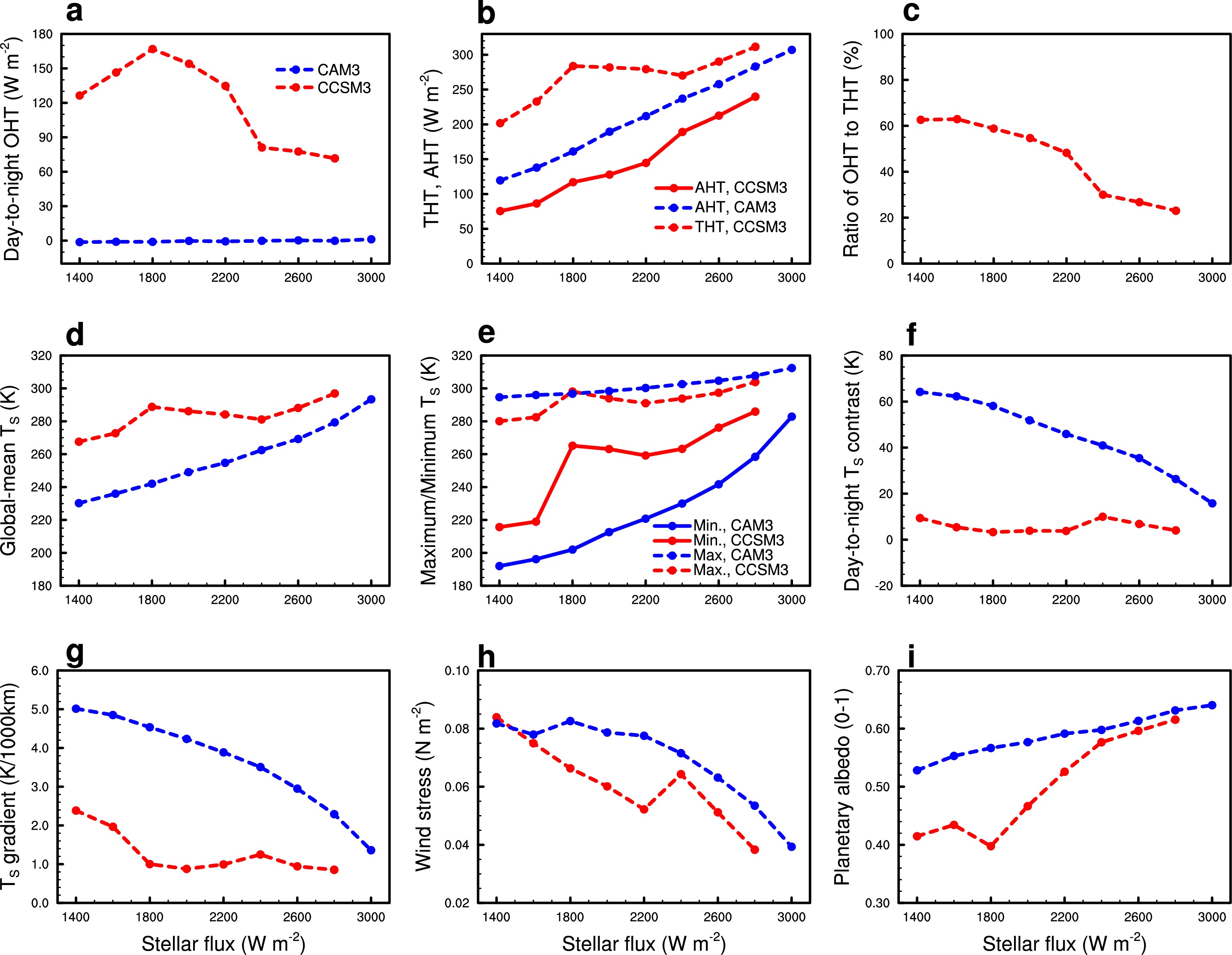}
\caption{The effect of ocean dynamics as a function of stellar flux. Blue line: the atmosphere-only 
model CAM3, and red line: the coupled ocean-atmosphere model CCSM3. (a), area-averaged day-to-night oceanic heat transport (OHT); (b), day-to-night atmospheric heat transport (AHT) and total heat transport (THT); (c), the ratio of OHT to THT in CCSM3; (d), global-mean surface temperature (T$_S$); (e) maximum surface temperature (dashed line) and minimum surface temperature (solid line); (f) day-to-night area-averaged T$_S$ contrast; (g) strength of surface temperature gradient ($\sqrt{(\frac{\partial T_S}{\partial x})^2 + (\frac{\partial T_S}{\partial y})^2} $ in units of K per 1,000 km);  (h) strength of surface wind stress ($\sqrt{\tau_x^2 + \tau_y^2}$ where $\tau_x$ is the zonal wind stress and $\tau_y$ is the meridional wind stress); and (i) planetary albedo, as a function of stellar flux. In all of the experiments, the rotation period is 60 Earth days and the stellar spectrum is a 4,500-K blackbody.}
\label{Figure:OHTAHTTHT}
\end{center}
\end{figure}

As the stellar flux is increased, the global-mean surface temperature 
(as well as the maximum and minimum surface temperatures) in CCSM3 generally 
increases, but the area-averaged 
day-to-night OHT decreases when the stellar flux is higher than 1,800~W\,m$^{-2}$ (Fig.~\ref{Figure:OHTAHTTHT}, 
for a discussion of the non-monotonic behavior of CCSM3's results, see section~\ref{nonlinear1800}). 
This finding seems counterintuitive because as the energy received by 
the day side of the planet increases, one might expect the ocean to transport 
more heat to the night side. This intuition does hold for the atmospheric and total 
(oceanic plus atmospheric) heat transports, which do increase with stellar 
flux (Fig.~\ref{Figure:OHTAHTTHT}(b)), but not for the OHT.

Fig.~\ref{Figure:OHTAHTTHT}(b) indicates that there is a compensation 
between OHT and atmospheric heat transport, but the compensation 
is imperfect. In CAM3, the OHT is zero and the atmospheric heat transport 
is higher than that in CCSM3. The total heat transport, however, is higher 
in CCSM3 than that in CAM3. The main reason is that the planetary 
albedo in CCSM3 is smaller (Fig.~\ref{Figure:OHTAHTTHT}(i)), meaning 
that more stellar radiation is absorbed by the dayside atmosphere and 
surface; therefore more energy can be supplied to transport by the 
atmosphere and ocean. This result is consistent with previous results 
\citep{Vallis:2009gu,EndertonandMarshall2013,FarnetiandVallis2013}: 
Increased atmospheric heat transport generally picks up the slack of 
reduced OHT, but the compensation is often not 100\,\%.

\begin{figure}[!htbp]
\begin{center}
\includegraphics[angle=0, width=16cm]{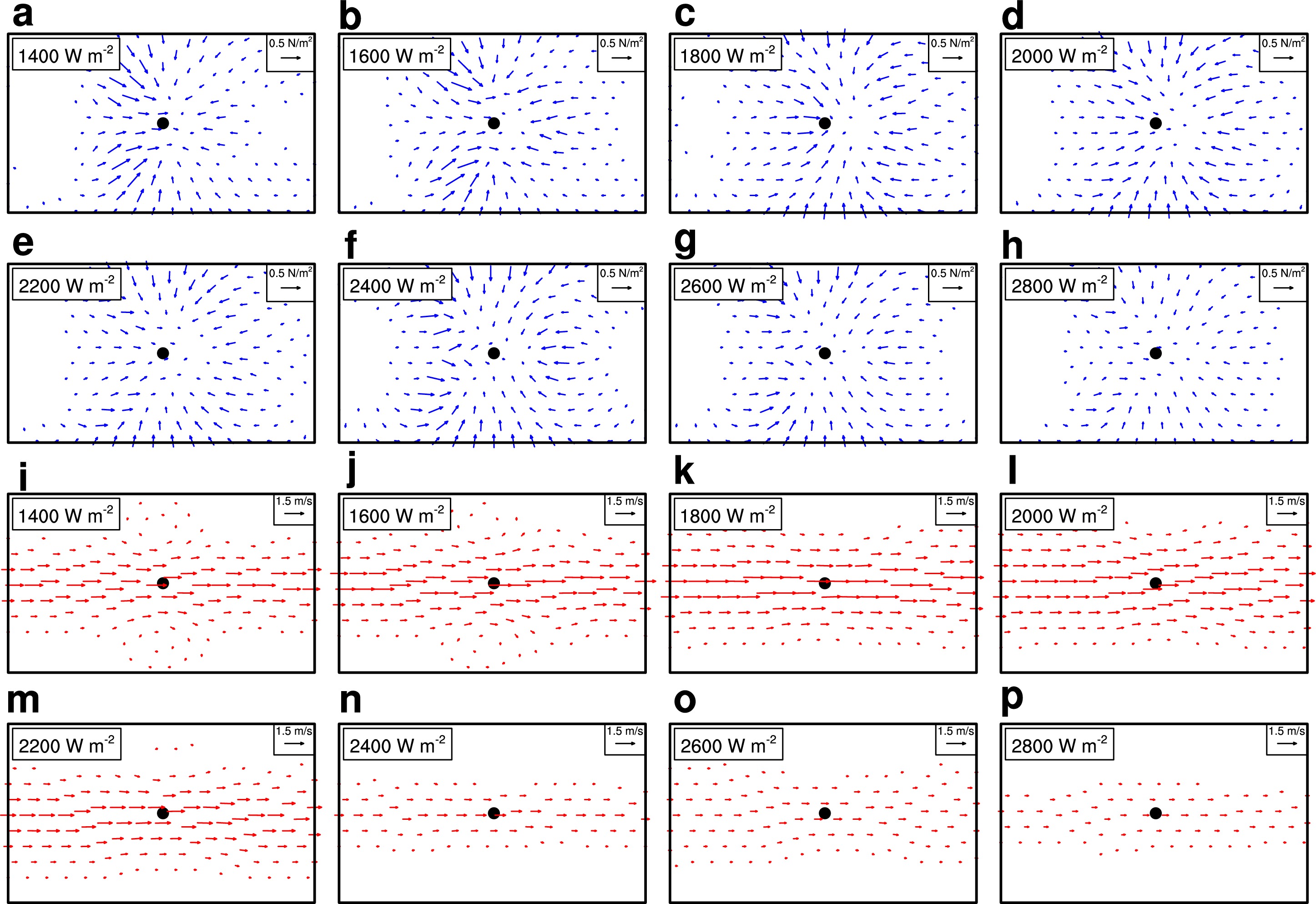}
\caption{The 1$^{st}$ reason for the weakening trend of ocean heat transport with increasing 
stellar flux: wind stresses weaken. 
Surface wind stresses on the ocean (a-f) and vertically averaged ocean currents (g-l) on 
tidally locked aqua-planets, as simulated by CCSM3. The stellar fluxes are 
1,400, 1,600, 1,800, 2,000, 2,200, 2,400, 2,600, and 2,800 W m$^{-2}$. The horizontal axis is longitude 
from 0$^{\circ}$ to 360$^{\circ}$, and the vertical axis is latitude from 85$^{\circ}$S to 85$^{\circ}$N. 
The black dot is the substellar point. The reference vector for stresses is 0.5 N m$^{-2}$, 
and values less than 0.05 N m$^{-2}$ are not plotted. The reference vector for ocean velocity 
is 1.5 m s$^{-1}$, and values less than 0.15 m s$^{-1}$ are not plotted.}
\label{Figure:OceanCurrents}
\end{center}
\end{figure}

The decreasing trend of OHT with increasing stellar flux results from 
weaker ocean currents combined with less stellar radiation depositing 
energy at the dayside sea surface. The main characteristic of ocean 
circulation on a tidally locked aqua-planet is west-east 
currents along the equator (Fig.~\ref{Figure:OceanCurrents}(i-p), ref.~\cite{hu2014role}). 
The ocean currents in the tropics result from the combined effect of surface wind 
stresses and equatorward momentum transport by large-scale Rossby and 
Kelvin waves in the ocean (more comprehensive 
analyses and detailed dynamical diagnostics will be addressed in a separate paper, 
Yaoxuan~Zeng, Yonggang Liu, \& Jun~Yang:~Understanding the wind-driven ocean circulation on 
a tidally locked aqua-planet, manuscript in preparation, 2018). Surface temperature gradients 
on the east side of the substellar point are smaller than those on its west side 
(see the right panel of Fig.~\ref{Figure:MiddleRange}(b)). 
Due to this asymmetry, the eastward stresses on the west side of the substellar 
point are generally stronger than the westward stresses on the east side of 
the substellar point (Fig.~\ref{Figure:OceanCurrents}(a-h)), so the net 
effect of the surface stresses is to drive eastward ocean flows. 
As the stellar flux is increased, the ocean flows become weaker 
(Fig.~\ref{Figure:OceanCurrents}(i-p)) due to weaker surface wind stresses (Figs.~\ref{Figure:OHTAHTTHT}(h) 
and~\ref{Figure:OceanCurrents}(a-h)). 
The decrease in surface wind stresses at least partly results from a smaller surface 
temperature gradient (Fig.~\ref{Figure:OHTAHTTHT}(g)), which is associated with the greater warming of 
the night side compared to the day 
side\footnote{Note that the surface temperature gradient in CCSM3 is much smaller than 
that in CAM3; this is due to the effect of ocean heat transport in the coupled model and 
associated feedback processes (see section~\ref{OHTMiddleRange}). Moreover, the 
surface temperature gradient is weak in all of CCSM3's experiments except for stellar fluxes 
of 1,400 and 1,600~W\,m$^{-2}$, but the surface wind stress generally 
decreases with increasing stellar flux (Fig.~\ref{Figure:OHTAHTTHT}(f--h)). 
This implies that surface temperature gradient 
is not the only determinant of the strength of surface wind stress; other factors, such 
as downward momentum flux from the free troposphere to the surface 
(chapter 12 of \cite{vallis2006atmospheric}), may be very important 
in certain conditions; and future work is required to fully understand this.}.

The ocean currents become ineffective when the stellar flux is 
equal to 2,400 W\,m$^{-2}$ or higher (Fig.~\ref{Figure:OceanCurrents}(n-p)).
Our result of small horizontal temperature gradients and weak surface wind 
stresses when the stellar flux is high is compatible with previous work on hot 
climates under high levels of insolation or atmospheric CO$_2$. Examples of this are 
shown in Fig.~1 of \cite{leconte2013increased} for the climate simulation of Earth 
under a brightening sun, in Fig.~2 of \cite{way2016Venus} for 
a temperate Venus, and in Fig.~4 of \cite{wolf2017assessing} 
for the exoplanet TRAPPIST 1d.

\begin{figure}[!htbp]
\begin{center}
\includegraphics[angle=0, width=16cm]{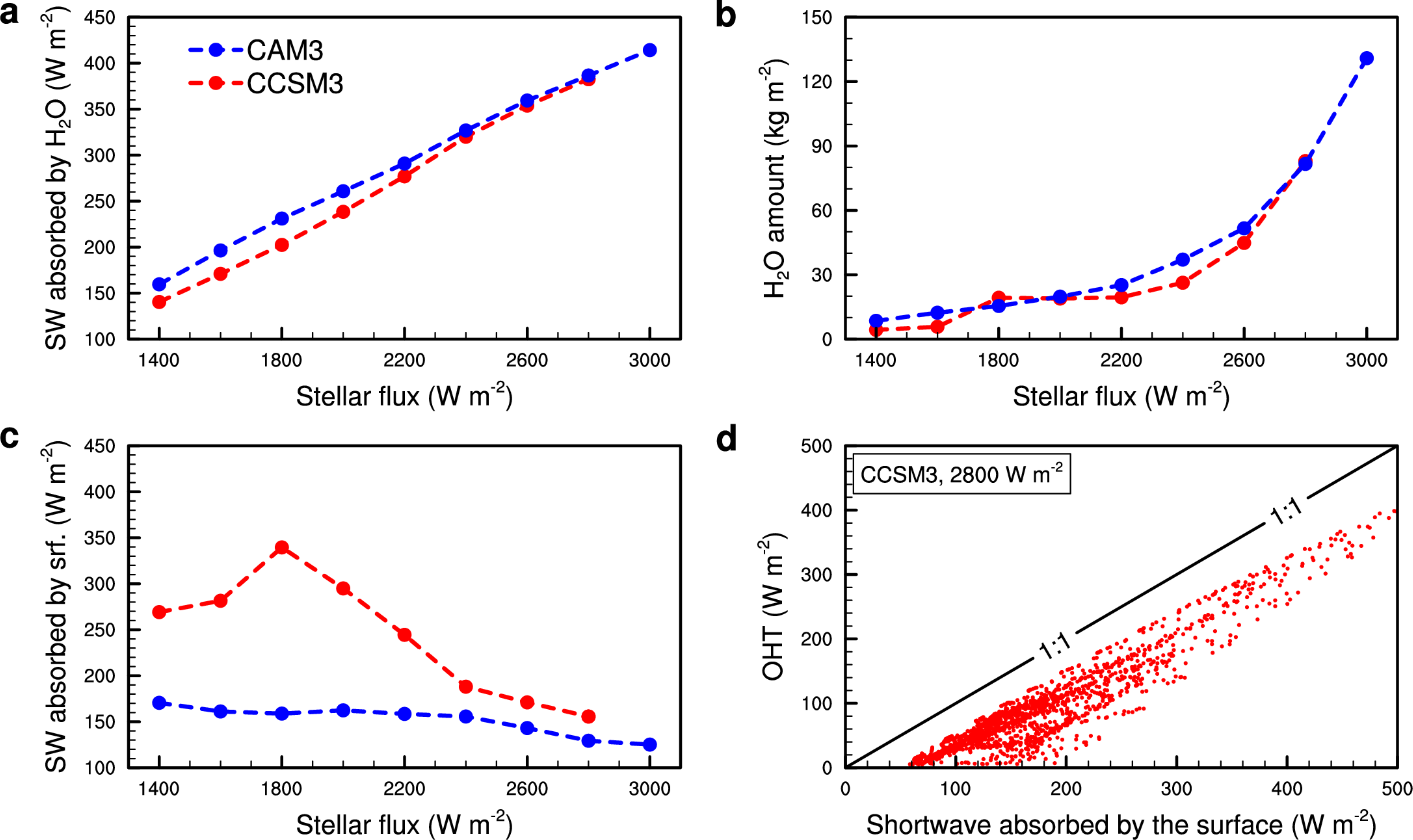}
\caption{The 2$^{nd}$ reason for the weakening trend of ocean heat transport with increasing 
stellar flux: surface stellar radiation decreases. 
Blue dot: CAM3; red dot: CCSM3. (a), shortwave (SW) radiation absorbed 
by water vapor on the day side; (b), vertically integrated water vapor amount on the day side; 
and (c), shortwave radiation absorbed by the surface (srf.) on the day side. (d), OHT vs 
shortwave absorbed by the surface at each grid cell of the model in CCSM3 with a stellar flux 
of 2,800 W\,m$^{-2}$; only data points that have positive values of OHT are shown. 
Note that OHT is smaller than the shortwave radiation absorbed by the surface for all data points; 
the cases of other stellar fluxes have the same relationship (figure not shown).}
\label{Figure:StellarAbsorption}
\end{center}
\end{figure}

As the stellar flux is increased, both shortwave absorption by water vapor 
(Fig.~\ref{Figure:StellarAbsorption}(a)) and shortwave reflection by clouds 
increase (Fig.~\ref{Figure:OHTAHTTHT}(i)). The increase in shortwave 
absorption by water vapor is primarily due to the increase in saturation 
vapor pressure with temperature following the Clausius-Clapeyron relation  
(Fig.~\ref{Figure:StellarAbsorption}(b)). The increase of shortwave reflection 
by clouds is due to the effect of a stabilizing cloud feedback: Greater stellar 
flux produces stronger substellar convection, more optically thick clouds, 
and a higher planetary albedo \citep{yang2013}. This phenomenon was first 
found in atmosphere-only general circulation models and briefly tested 
in CCSM3; here we confirm that this feedback exists in a broader 
range of coupled ocean-atmosphere simulations. Both the increased 
water vapor absorption and enhanced cloud reflection lead to the surprising 
result that the stellar energy reaching the sea surface actually decreases 
as the stellar flux at the top of the atmosphere is 
increased (Fig.~\ref{Figure:StellarAbsorption}(c)).

For long-time mean climatology, the surface energy balance of an ice-free 
region can be written as: OHT\,$=$\,R$_{s}$\,$-$\,R$_{l}$\,$-$\,SH\,$-$\,LH, 
where R$_{s}$ is net shortwave radiation flux into the surface, R$_{l}$ is net longwave 
radiation flux from the surface to the atmosphere, SH is sensible heat flux from 
the surface to the atmosphere, and LH is latent heat flux from the surface to 
the atmosphere. All of these variables 
are positive as long as there is no temperature inversion (under an inversion, SH 
and R$_{l}$ will be negative). This equation implies that OHT $<$ R$_{s}$, 
i.e., the maximum allowed OHT is constrained by the absorbed shortwave 
radiation at the sea surface, as shown in Fig.~\ref{Figure:StellarAbsorption}(d). 
The value of R$_{s}$ decreases with increasing stellar flux 
(Fig.~\ref{Figure:StellarAbsorption}(c)), so that OHT should have the same trend.

\begin{figure}[!htbp]
\begin{center}
\includegraphics[angle=0, width=14cm]{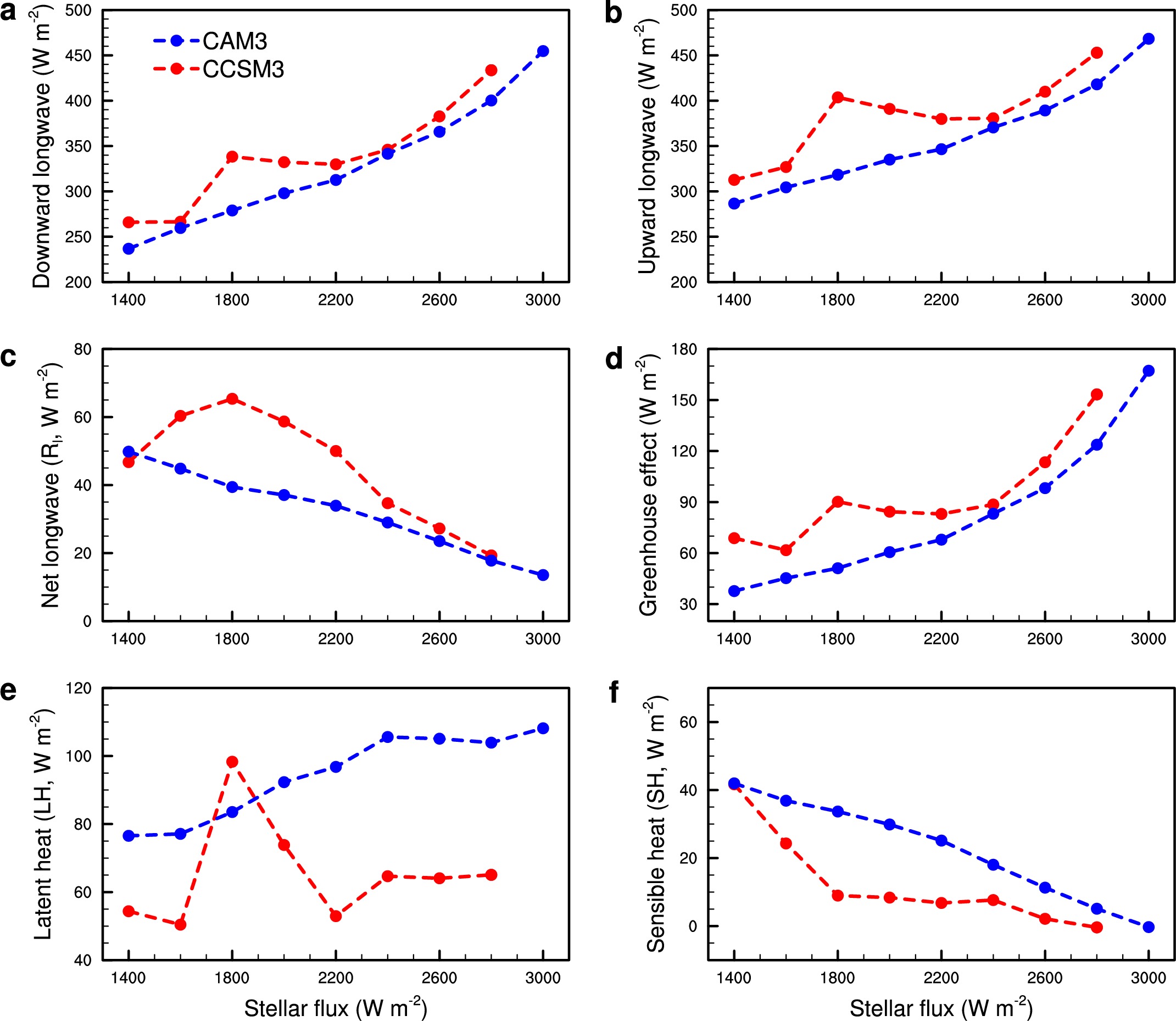}
\caption{Surface energy flux and greenhouse effect as a function of stellar flux. 
(a) downward longwave radiation flux at the surface, (b) upward longwave radiation 
flux at the surface, (c) net longwave radiation flux (upward minus downward), 
(d) clear-sky greenhouse effect, defined as $\sigma T_s^4 - OLR_{clear}$ where 
$T_s$ is the surface temperature, OLR$_{clear}$ is the clear-sky outgoing longwave 
radiation at the top of the model, and $\sigma$ is the Stefan-Boltzmann 
constant (ref. \cite{Pierrehumbert2005}), 
(e) latent heat flux at the surface, and (f) sensible heat flux at the surface. 
All of these variables are area averages on the day side only.}
\label{Figure:EnergyBalance}
\end{center}
\end{figure}

Note that although the value of R$_{s}$ decreases with stellar flux, 
the surface temperatures generally keep increasing (Fig.~\ref{Figure:OHTAHTTHT}(d--e)); 
this is mainly due to the increasing of atmospheric greenhouse effect. The physical process 
can be briefly summarized follows: when the stellar flux is increased, more shortwave radiation 
is absorbed by water vapor (Fig.~\ref{Figure:StellarAbsorption}(a)), which acts to 
increase the air temperature and thereby the atmosphere is able to hold more 
water vapor based on the Clausius-Clapeyron relation. The increased water vapor 
concentration raises the greenhouse effect of the atmosphere 
(Fig.~\ref{Figure:EnergyBalance}(d)) and emits more longwave radiation to the 
surface (Fig.~\ref{Figure:EnergyBalance}(a)) such that the net longwave radiation 
at the surface decreases with stellar flux (Fig.~\ref{Figure:EnergyBalance}(c)), warming the surface. 
Moreover, as the stellar flux is increased, the surface sensible heat flux generally 
decreases (Fig.~\ref{Figure:EnergyBalance}(f)), which has a secondary warming 
effect on the surface.

The day-to-night OHT decreases with increasing stellar flux because of the 
above two mechanisms, weaker surface wind stresses and less stellar 
energy deposited at the surface. In the 1,400 W\,m$^{-2}$ simulation, the OHT 
is 124 W\,m$^{-2}$, which contributes to 63\,\% of the total heat transport, 
whereas in the 2,800 W\,m$^{-2}$ simulation, it decreases to 70~W\,m$^{-2}$ 
and the percentage reduces to 23\,\% (Fig.~\ref{Figure:OHTAHTTHT}(c)). 
These results indicate 
that at the inner edge of the habitable zone, the OHT would be even smaller, although 
the total (atmosphere plus ocean) heat transport would still be very robust.

\begin{figure}[!htbp]
\begin{center}
\includegraphics[angle=0, width=16cm]{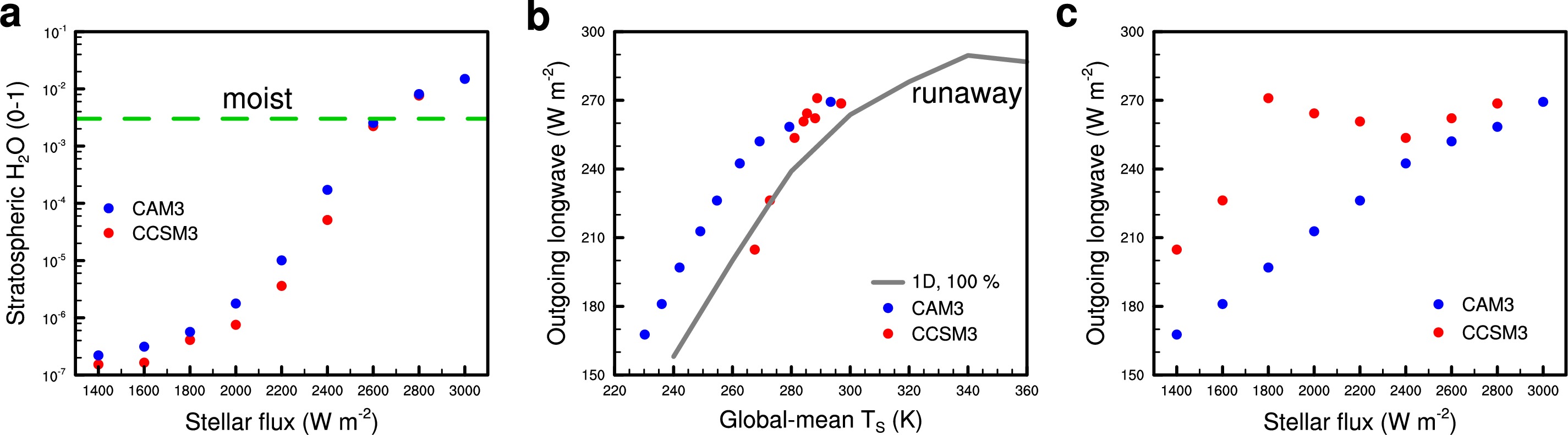}
\caption{The effect of ocean dynamics on the inner edge of the habitable zone 
for a tidally locked aqua-planet around a low-mass star. 
Red dots: CCSM3, and blue dots: CAM3. (a), global-mean stratospheric water vapor 
mixing ratio at 50 hPa, as a function of stellar flux (horizontal green line: the moist greenhouse 
limit of 3,000 ppmv). (b), global-mean outgoing longwave radiation (OLR) as a function 
of surface temperature. The gray line in (b) is the OLR from a 1D cloud-free radiative transfer model 
assuming 100\,\% relative humidity, as calculated by \cite{leconte2013increased}. 
(c) OLR as a function of stellar flux.}
\label{Figure:InnerEdge}
\end{center}
\end{figure}

The decreasing trend of OHT with increasing stellar flux suggests that ocean 
dynamics may be not important for very hot climates. Indeed, we find that 
the location of the inner edge of the habitable zone does not depend on 
ocean dynamics (Fig.~\ref{Figure:InnerEdge}). For a stellar temperature of 4,500 K, the climate 
system enters a moist greenhouse state at a stellar flux of $\simeq$\,2,600--2,800 W\,m$^{-2}$ for 
both CCSM3 and CAM3 (Fig.~\ref{Figure:InnerEdge}(a)). At this level of stellar flux, the 
stratospheric water vapor concentration, $\ge$\,$3,000$ ppmv, is high enough that 
H$_2$O photo-dissociation and subsequent H escape to space becomes significant, 
which is known as the moist greenhouse state \citep{Kasting88,Kasting93}. It should be 
mentioned that at a stellar flux of 2,600 W\,m$^{-2}$, 
the day-to-night OHT is still about 25\,\% of the total heat transport and the difference 
in global-mean surface temperature between CCSM3 and CAM3 is still up to 25~K. 
However, the key variable for the onset of the moist greenhouse state is the stratospheric 
water vapor concentration rather than the global-mean surface temperature or other variable(s).

For lower stellar fluxes, stratospheric water vapor in CCSM3 is slightly lower than 
in CAM3 (Fig.~\ref{Figure:InnerEdge}(a)), due to relatively weaker vertical velocities 
in the stratosphere over the substellar region in CCSM3 than those in CAM3 (figure 
not shown). At higher stellar flux, however, the day-to-night OHT become weaker 
and the two models 
have the same concentration of stratospheric water vapor. On tidally locked planets, 
the stratospheric water vapor abundance is primarily determined by the temperature 
of the tropopause and the strength of 
the stratospheric vertical velocity above the substellar region. The vertical velocity 
is mainly driven by near-infrared radiative heating associated with shortwave 
absorption by stratospheric water vapor and cloud particles \citep{Fujiietal2017:Moist}.

Finally, our results suggest that the ocean-atmosphere and atmosphere-only models would 
enter into a runaway greenhouse state, 
in which absorbed shortwave exceeds maximum allowed outgoing longwave ($OLR_{max}$), 
at a similar stellar flux. 
The stellar flux limit (S$_{limit}$) for triggering the runaway greenhouse only depends on 
$OLR_{max}$ and the planetary albedo 
near the inner edge ($\alpha_p$), i.e., S$_{limit}$ = $4\times OLR_{max}/(1-\alpha_p) $, 
where the factor 4 is the ratio of a sphere’s surface area to its cross-sectional area. 
Figure~\ref{Figure:InnerEdge}(b) shows that the two 
models likely have the same $OLR_{max}$. 
Meanwhile, they exhibit nearly the same $\alpha_p$ as the stellar flux is 
increased to 2,800 W m$^{-2}$ (Fig.~\ref{Figure:OHTAHTTHT}(i)), 
although the global-mean surface temperature still has a difference of 20 K and 
the day-to-night OHT is still 23\,\% of the total heat transport. 
In the models we used, the stellar flux limit\footnote{There is a significant 
uncertainty in $OLR_{max}$ (as well as in 
$\alpha_p$), 295 $\pm$ 15 W m$^{-2}$, arising from clouds, the degree of 
atmospheric sub-saturation and the uncertainties in radiative transfer 
calculations (\cite{leconte2013increased,wolf2015evolution,Yang2016,Marcqetal2017}). 
For a star with a temperature of 4,500 K, the value of $\alpha_p$ is $\simeq$\,0.62 
(Fig.~\ref{Figure:OHTAHTTHT}f), and therefore the stellar flux limit is about 
3,100\,$\pm$\,160 W\,m$^{-2}$. If we further assume a 10\,\% uncertainty (it could be 
even larger because clouds are not explicitly resolved and models employ 
different cloud parameterization schemes) 
in $\alpha_p$, the stellar flux limit would be about 3,100\,$\pm$\,650 W\,m$^{-2}$. 
For the models used in this study, the maximum allowed clear-sky outgoing 
longwave radiation is about 355 W\,m$^{-2}$ (see Fig.~7(c) in \cite{Yangetal2018ExoMIP}), 
the cloud longwave radiative effect is about 45 W m$^{-2}$ (Fig.~\ref{Figure:CloudRE}(a)) 
and the value of $\alpha_p$ is about 0.62, so that the runaway greenhouse limit is 
about 3,260~W\,m$^{-2}$. Our last converged experiment, 2,800~W\,m$^{-2}$, is 
still 460~W\,m$^{-2}$ less than this limit.} is about 3,300 W\,m$^{-2}$ 
both with and without ocean dynamics. 
This limit is about two times that for rapidly rotating planets around G stars 
(such as Earth, \cite{leconte2013increased}), 
primarily due to the stabilizing cloud feedback \citep{yang2013}.
In summary, ocean dynamics do not influence the stellar flux limit for the onset 
of the runaway greenhouse state in our experiments using a 60-day 
tidally locked orbit.

As shown in Fig.~\ref{Figure:InnerEdge}(b), for lower stellar fluxes and under 
the same global-mean surface temperature, outgoing longwave radiation 
at the top of the model in CAM3 is higher than in CCSM3. This is due 
to the absence of a nightside temperature inversion (Fig.~\ref{Figure:MiddleRange}(f)) 
and the stronger longwave cloud 
radiative effect in CCSM3 (Fig.~\ref{Figure:CloudRE}), both 
of which reduce the outgoing longwave radiation in the coupled 
atmosphere-ocean model. When the stellar flux is $\ge$ 3,000 W\,m$^{-2}$, 
the temperature inversion in CAM3 disappears
and the longwave cloud radiative effect becomes close to that of CCSM3 
(Fig.~\ref{Figure:CloudRE}(b)), so that the outgoing longwave 
radiation fluxes of the two models coverge to each other (Fig.~\ref{Figure:InnerEdge}(b)). 
Note that due to numerical instability, CCSM3 blew up 
for stellar fluxes higher than 2,800 W\,m$^{-2}$, so that the 
maximum allowed outgoing longwave radiation can not be read from Fig.~\ref{Figure:InnerEdge}(b), 
however, our present experiments show no evidence that ocean dynamics 
affects the maximum allowed outgoing longwave radiation and 
the planetary albedo at the inner edge of the habitable zone. 

Another way to understand the differences and similarities between 
CAM3 and CCSM3 is plotting the global-mean outgoing longwave radiation 
as a function of the stellar radiation downward at the top of the atmosphere, 
as shown in Fig.~\ref{Figure:InnerEdge}(c). At equilibrium, outgoing longwave 
radiation is equal to absorbed stellar radiation. At lower stellar fluxes, 
the outgoing longwave radiation is higher in CCSM3 than in  CAM3 
because CCSM3 has a lower planetary albedo 
(see Fig.~\ref{Figure:OHTAHTTHT}(i)) and therefore absorbs more stellar radiation. 
For a high stellar flux, $\ge$2,400 W\,m$^{-2}$, the difference in the 
outgoing longwave radiation between CAM3 and CCSM3 becomes small; 
this again suggests that ocean dynamics likely have no 
significant effect on the inner edge of the habitable zone.

\subsection{Observational Thermal Phase Curves}
\label{PhaseCurves}

\begin{figure}[!htbp]
\begin{center}
\includegraphics[angle=0, width=15cm]{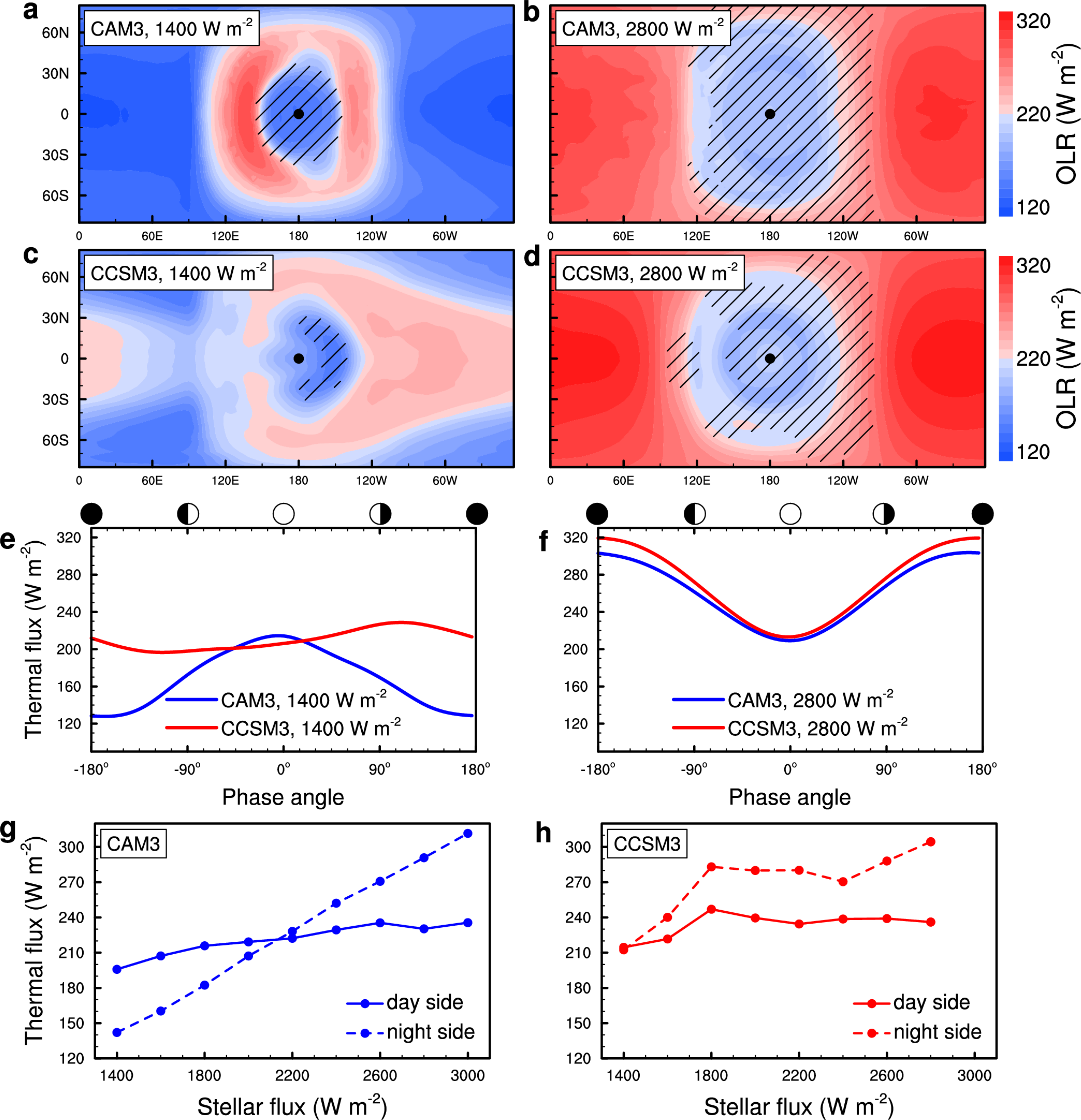}
\caption{Effects of ocean dynamics on thermal phase curves of planets in the middle 
range (left columns) and near the inner edge (right columns) of the habitable zone. 
(a-d), outgoing longwave radiation (OLR, color shading) and vertically integrated cloud 
water content higher than 400 g m$^{-2}$ (masked by oblique lines). (a), CAM3’s simulation 
with a stellar flux of 1,400 W m$^{-2}$; (b), CAM3 and 2,800 W m$^{-2}$; (c), CCSM3 
and 1,400 W m$^{-2}$; and (d), CCSM3 and 2,800 W m$^{-2}$. (e), thermal phase 
curves of CAM3 (blue line) and CCSM3 (red line) with a stellar flux of 1,400 W m$^{-2}$. 
(f), same as (e) but for 2,800 W m$^{-2}$. The observer views the day side of the planet 
at a phase angle of 0$^{\circ}$ and sees the night side at phase angles of $\pm$180$^{\circ}$. 
(g), time-mean thermal flux on the day side (solid line) and the night side (dashed line) in CAM3. 
(h), same as (g) but for CCSM3. Note in (a-d), due to the effect of clouds and water vapor, 
there is a decrease in OLR over the substellar point in all experiments.}
\label{Figure:PhaseCurves}
\end{center}
\end{figure}

We further find that ocean dynamics have a very small effect on the observational 
thermal phase curve of tidally locked planets near the inner edge of the habitable zone. 
Figure~\ref{Figure:PhaseCurves} shows the spatial pattern of thermal emission at 
the top of the atmosphere and thermal phase curves. The phase curves are 
disk-integrated thermal radiation measured by an observer as a function of the 
planet’s position in its orbit \citep[e.g.,][]{CowanandAgol2008,koll-2015}. 
The curves are determined by the combined effect of surface temperature, 
water vapor, clouds, and atmospheric and oceanic heat transports; therefore, 
they can be used to probe the atmosphere and/or surface characteristics of 
exoplanets.

For planets near the inner edge of the habitable zone, the day-to-night 
OHT is not strong, so that the phase curves of CAM3 and CCSM3 are similar 
(Fig.~\ref{Figure:PhaseCurves}(f)). Importantly, the thermal emission on the night 
side is much higher than that on the day side, reversing the day-night thermal 
contrast (Fig.~\ref{Figure:PhaseCurves}(b,d)). This reversal is due to the 
high concentration of water vapor and clouds above the substellar region, which 
absorb the thermal radiation from the surface but re-emit to space at much 
lower temperatures \citep{yang2013,haqq2017demarcating}. Meanwhile, the 
night side is relatively dryer and therefore infrared radiation from near the surface can be lost to 
space easily \citep{yang2014-low,Pierrehumbert-1995:thermostats}. In CAM3, 
the phase curve reversal occurs when the stellar flux is equal to or higher than 
2,200~W\,m$^{-2}$ whereas in CCSM3 it occurs at a much lower stellar flux,  
1,600~W\,m$^{-2}$ (Fig.~\ref{Figure:PhaseCurves}(g,h)). This suggests that planets 
with deep oceans in the middle range of the habitable zone could also have 
higher thermal emission on the night side than on the day side.

For a stellar flux of 1,400~W\,m$^{-2}$ in CAM3, the phase curve has a maximum 
when the observer sees the day side and a minimum when the observer views 
the night side (Fig.~\ref{Figure:PhaseCurves}(e)). When ocean dynamics are 
included, more heat is transported to the night side, so that the day-night 
thermal contrast in CCSM3 is much smaller than that in CAM3 and therefore  
the amplitude of the phase curve is much smaller. Furthermore, because 
most of the heat is transported to the east side (rather than the west side) 
of the substellar point (Fig.~\ref{Figure:PhaseCurves}(c)), the ridge of the 
phase curve exhibits a positive phase angle displacement of $\simeq$\,120$^{\circ}$ 
(Fig.~\ref{Figure:PhaseCurves}(e)). In contrast, when ocean dynamics are not considered, the thermal emission 
on the west side of the substellar point is relatively higher than 
that on the east side (Fig.~\ref{Figure:PhaseCurves}(a)), because water 
vapor is transported eastward, where it absorbs thermal emission from the surface, 
and therefore the thermal phase curve exhibits a small negative 
phase angle displacement (Fig.~\ref{Figure:PhaseCurves}(e), 
see also Fig.~3(b) in \cite{yang2013}).


Could the oceanic effect on the thermal phase curve be observed by 
the \textsl{James Webb Space Telescope (JWST)}?  
To estimate JWST's precision, we assume that observations are only limited by 
photon noise and by the telescope's detector efficiency (see \cite{koll-2015}). 
For the target we assume the LHS 1140 system \citep{Dittmannetal2017:LHS1140b}; 
closer or brighter host stars would be easier to measure. For the instrument we assume 
MIRI-F1800W (16.5-19.5 micron) photometry and a photon efficiency of $1/3$. 
Because habitable zone 
planets are cool, it is necessary to observe at long wavelengths to obtain favorable planet-star thermal 
contrasts. We note that at such long wavelengths additional sources of error could become significant, 
e.g., dust or thermal background, so our estimate is optimistic. 
For a 2-hour integration the 1 sigma error bar for the flux will be 144~W\,m$^{-2}$. 
For a 24-hour integration, the error goes down by $1/\sqrt{12}$, so a 1 sigma error would be $\simeq$42~W\,m$^{-2}$, larger than the amplitude of the thermal phase curve shift under ocean dynamics 
in the middle of the habitable zone (see red line in Fig.~\ref{Figure:PhaseCurves}(e)) but 
smaller than the amplitude of the day-night phase curve reversal near the inner edge of the habitable zone  (Fig.~\ref{Figure:PhaseCurves}(f)). Therefore, JWST observations with long 
staring exposures would be able to detect the thermal phase curve 
reversal near the inner edge of the habitable zone. However, unless future discoveries detect a 
target more favorable than LHS 1140b, the ocean-induced thermal phase shift would likely not 
detectable with JWST.



\subsection{Non-monotonic Behavior of the Coupled Atmosphere--Ocean System}
\label{nonlinear1800}

As shown in Figs.~\ref{Figure:CloudRE}, \ref{Figure:OHTAHTTHT}, \ref{Figure:StellarAbsorption}
and \ref{Figure:EnergyBalance}, the climate in CCSM3 is not 
a monotonic function of stellar flux. Variables that show the non-monotonic behavior include: 
(1) the day-to-night ocean heat transport (Fig.~\ref{Figure:OHTAHTTHT}(a)), night-side 
longwave cloud radiative effect (Fig.~\ref{Figure:CloudRE}(b)), day-side shortwave absorption 
by the sea surface (Fig.~\ref{Figure:StellarAbsorption}(c)), and surface net longwave radiation 
(Fig.~\ref{Figure:EnergyBalance}(c)) increase with stellar flux 
between 1,400 and 1,800~W\,m$^{-2}$ but decrease with stellar flux when it is higher than 
1,800~W\,m$^{-2}$; (2) the global-mean surface temperature (Fig.~\ref{Figure:OHTAHTTHT}(d)),  
day-to-night total heat transport (Fig.~\ref{Figure:OHTAHTTHT}(b)), global-mean longwave 
cloud radiative effect (Fig.~\ref{Figure:CloudRE}(a)), and both upward and downward longwave 
radiation fluxes at the surface (Fig.~\ref{Figure:EnergyBalance}(a \& b)) increase between 
1,400 and 1,800~W\,m$^{-2}$, decrease between 1,800 and 2,400~W\,m$^{-2}$, and 
increase again when the stellar flux is higher than 2,400~W\,m$^{-2}$; 
(3) the planetary albedo (Fig.~\ref{Figure:OHTAHTTHT}(i))  
mostly increases with stellar flux but has a minimum value when the stellar flux is 1,800~W\,m$^{-2}$.
Sensitivity tests show that this non-monotonic behavior seems don't depend on the initial state (Fig.~\ref{Figure:TimeSeries}(e-f)). 

\begin{figure}[!htbp]
\begin{center}
\includegraphics[angle=0, width=15cm]{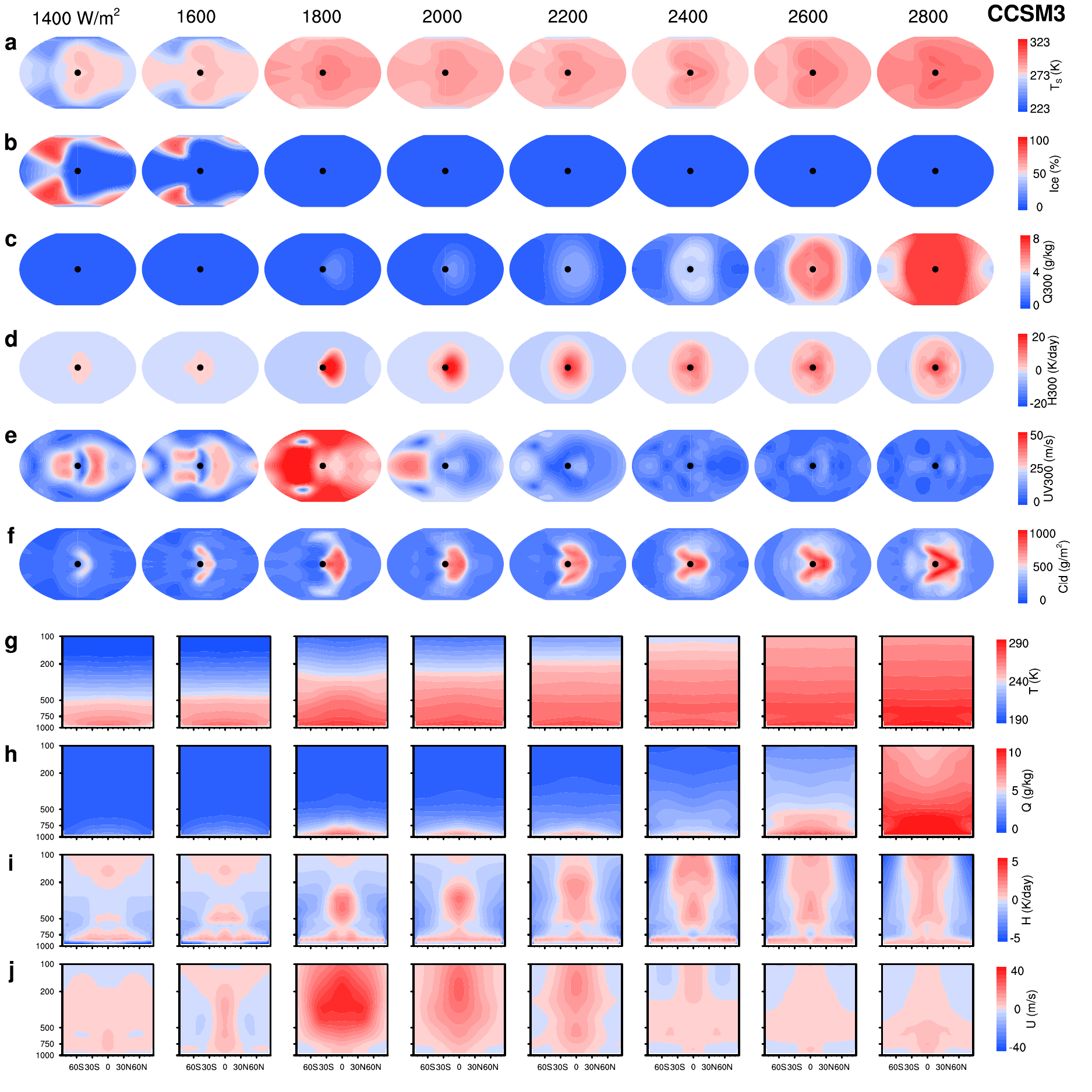}
\caption{Atmospheric characteristics simulated by CCSM3 at stellar fluxes 
from 1,400 to 2,800~W\,m$^{-2}$ with an interval of 200~W\,m$^{-2}$. 
(a) Surface air temperature, (b) sea ice coverage, (c) water vapor specific humidity at 300 hPa, 
(d) heating rate at 300 hPa due to the combined effect of shortwave heating, longwave cooling, 
and moist processes (such as latent heat release during condensation), (e) horizontal wind 
strength ($\sqrt{u^2+v^2}$) at 300 hPa, (f) vertically integrated cloud water amount, (g) 
zonal-mean air temperature, (h) zonal-mean water vapor specific humidity, (i) zonal-mean heating 
rate, and (j) zonal-mean zonal winds. Note that the troposphere and water vapor profiles extend 
to higher altitudes in a warmer climate, so that more stellar radiation is absorbed by water 
vapor at high altitudes. The black dot in (a--f) denotes the substellar point.}
\label{Figure:CCSM3nonlinearity}
\end{center}
\end{figure}

\begin{figure}[!htbp]
\begin{center}
\includegraphics[angle=0, width=15cm]{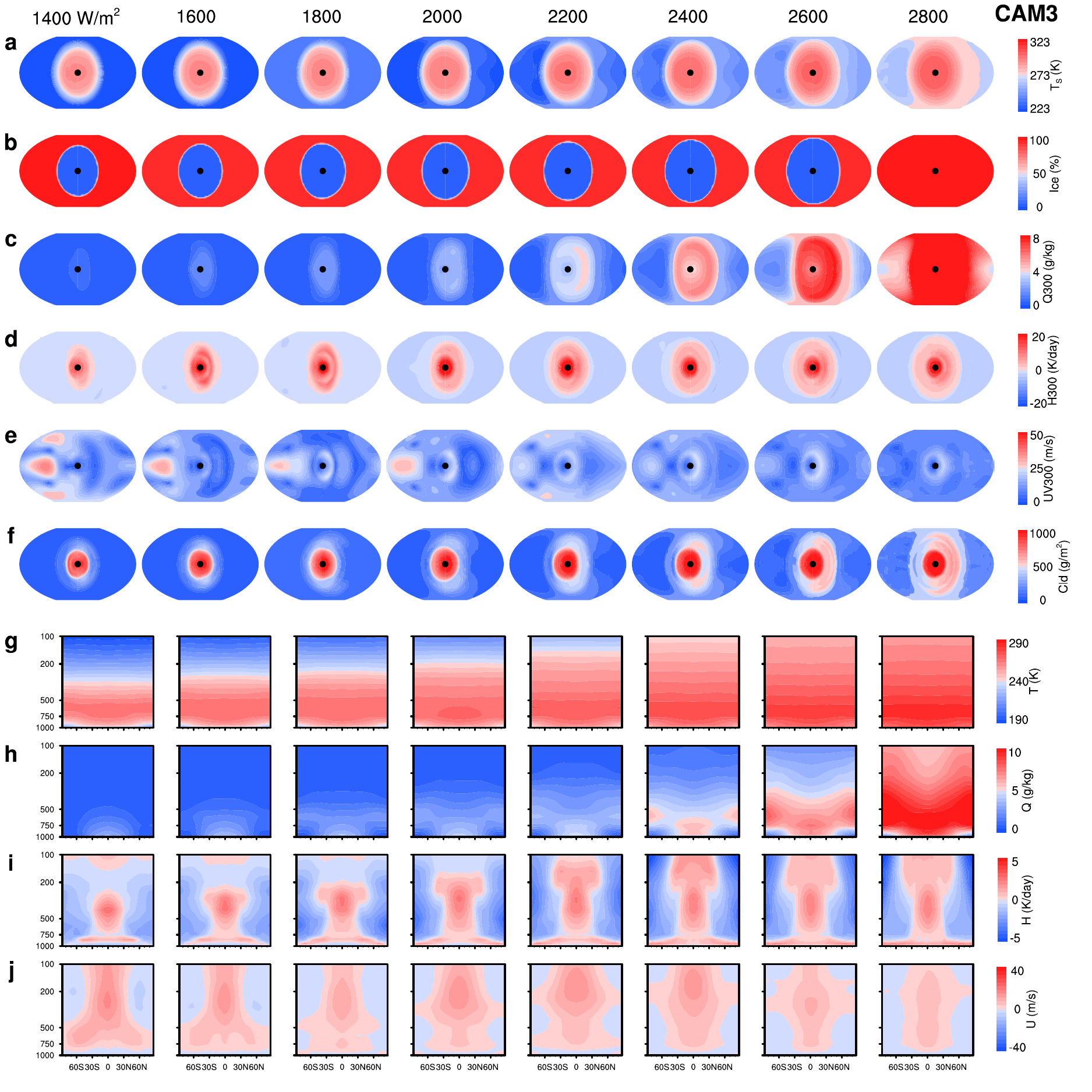}
\caption{Same as Fig.~\ref{Figure:CCSM3nonlinearity} but CAM3 simulations.}
\label{Figure:CAM3linearity}
\end{center}
\end{figure}

The non-monotonic behavior in CCSM3 likely results from atmosphere--ocean 
interactions and associated feedback processes because the climate simulated using 
the atmosphere-only model CAM3 is close to monotonic (see Figs.~\ref{Figure:CloudRE}, ~\ref{Figure:OHTAHTTHT}, \ref{Figure:StellarAbsorption}, \& \ref{Figure:EnergyBalance}). 
The increase in global-mean surface temperature between 1,400 and 1,600~W\,m$^{-2}$ and 
between 2,400 and 2,800~W\,m$^{-2}$ in CCSM3 is relatively easier to understand, whereas  
the slight decrease in global-mean surface temperature between 1,800 and 2,200~W\,m$^{-2}$ 
and the minimum in planetary albedo at 1,800~W\,m$^{-2}$ are 
due to the complex interactions between the ocean, atmosphere and clouds.
Comparing the 1,400 and 1,600~W\,m$^{-2}$ cases, the night side is covered by sea ice in both experiments 
(Fig.~\ref{Figure:CCSM3nonlinearity}(b)) and the surface temperature gradients decrease 
but not very significantly (Figs.~\ref{Figure:CCSM3nonlinearity}(a) \& \ref{Figure:OHTAHTTHT}(f-g)), 
so that the OHT increases with increasing stellar flux. This implies  
that in these two experiments of relatively lower stellar flux the two mechanisms---the surface temperature gradient decreasing and the surface wind stress weakening addressed in the section~\ref{OHTtrend} are not active enough to counteract the effect of increasing stellar flux. In the 2,400, 2,600 and 2,800~W\,m$^{-2}$ experiments, the two mechanisms are very 
effective in reducing the day-to-night OHT although the stellar flux is increased 
and the atmospheric heat transport increases with stellar flux; as shown in 
Fig.~\ref{Figure:OceanCurrents}, the ocean currents in these three cases 
are much weaker than those in all other experiments.
For the cases between 1,800 and 2,200~W\,m$^{-2}$, the situation is more complex and 
the key may be associated with the effect of ocean dynamics on the spatial pattern of sea 
surface temperature and consequently on the atmospheric heating rate, the 
strength of atmospheric superrotation (winds blowing from west to east 
over the deep-tropical region) and the spatial pattern of clouds 
(Fig.~\ref{Figure:CCSM3nonlinearity}). In the 1,800~W\,m$^{-2}$ case of CCSM3, 
strong ocean currents transport relatively cold seawater from the night side to the 
west side of the substellar point and also transport relatively warm 
seawater from the substellar point to its east side (similar behavior also occurs at 2,000 and 
2,200~W\,m$^{-2}$, but the 1,800~W\,m$^{-2}$ case is the most significant). As a result, 
the sea surface temperature exhibits a strong, zonally asymmetric 
pattern with the east side of the substellar point much warmer than 
the west side (Fig.~\ref{Figure:CCSM3nonlinearity}(a)). This asymmetric pattern 
causes the water vapor concentration and atmospheric heating rate to 
show a similar zonally asymmetric pattern (Fig.~\ref{Figure:CCSM3nonlinearity}(c-d)). 
The asymmetric heating pattern is likely more effective at generating 
Rossby waves and Kelvin waves in the atmosphere; the waves pump eastward 
momentum from higher latitudes to the equator \citep{ShowmanandPolvani2011}, 
inducing very strong super-rotating winds over the substellar region 
(Fig.~\ref{Figure:CCSM3nonlinearity}(e \& j)); these winds advect clouds eastward away from 
the substellar region where the stellar flux peaks, and the 
west side of the substellar point becomes nearly cloud-free 
(Fig.~\ref{Figure:CCSM3nonlinearity}(f)); as a result, the planetary albedo becomes 
smaller (0.39 in the case of 1,800~W\,m$^{-2}$ versus 0.44 in the case of 1,600~W\,m$^{-2}$, suggesting that 
the stabilizing cloud feedback is less active for this case); therefore, more 
stellar radiation reaches the surface, warming the sea surface. In CAM3, this 
phenomenon does not occur because the day-side sea surface temperature is nearly 
symmetric around the substellar point and the atmospheric zonal winds are weak 
in all of the experiments (Fig.~\ref{Figure:CAM3linearity}).

Moreover, in CCSM3 the strength of the atmospheric super-rotation 
decreases with stellar flux between 1,800 and 2,200~W\,m$^{-2}$ 
(Fig.~\ref{Figure:CCSM3nonlinearity}(j)), so that the planetary albedo increases with stellar flux,  
which enhances the primary effect of the stabilizing cloud feedback with increasing stellar flux \citep{yang2013}.  
The planetary albedo increases with stellar flux fastest in these three experiments  (Fig.~\ref{Figure:OHTAHTTHT}(i)) and as a result the global-mean surface temperature decreases sightly (rather than increases) with stellar flux (Fig.~\ref{Figure:OHTAHTTHT}(d)). 
The decrease in OHT of these three experiments is due to the combined effect of the planetary albedo 
increasing (so less surface shortwave radiation is deposited) and surface wind stress weakening (Figs.~\ref{Figure:OHTAHTTHT}(h) \& \ref{Figure:OceanCurrents}). Further work is required to 
clearly understand the onset condition for strong atmospheric super-rotation in the coupled model, especially as relates to the role of the spatial pattern of surface temperature or atmospheric heating.

Overall, the non-monotonicity of CCSM3 is likely due to complex 
interactions among the sea surface, atmospheric circulation, water vapor and clouds. This further 
demonstrates that why fully coupled atmosphere-ocean modeling is required to simulate the details 
of the climate of planets in the middle of the habitable zone. 

\section{Discussion}
\label{sec:discussion}

\begin{figure}[!htbp]
\begin{center}
\includegraphics[angle=0, width=15cm]{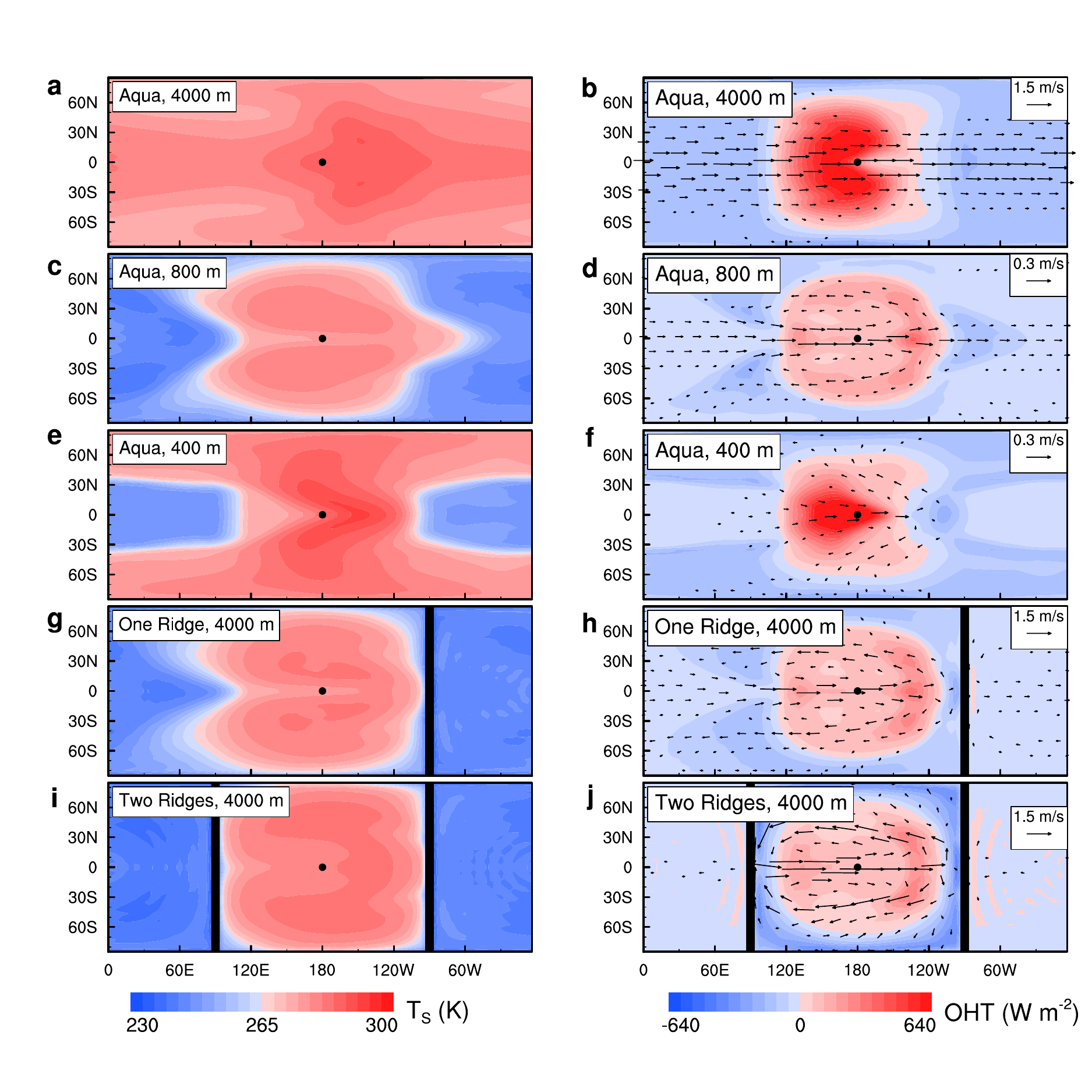}
\caption{Effects of ocean depth and continental barriers on the long-time mean surface 
temperature (left panels) and ocean heat transport (OHT, right panels, negative: heat from 
ocean to atmosphere, positive: heat from atmosphere to ocean). (a \& b), an aqua-planet 
with an uniform ocean depth of 4,000 m; (c \& d), an ocean depth of 800 m; (e \& f), an ocean 
depth of 400 m; (g \& h), one-ridge world with an ocean depth of 4,000 m; and (i \& j), 
two-ridges world with an ocean depth of 4,000 m. In the right panels, the vectors denote 
vertically averaged ocean currents with reference vectors of 1.5, 0.3, 0.3, 1.5, and 
1.5 m\,s$^{-1}$, respectively. The thick black lines in (g-j) denote the 
continental barriers, which extend from the ocean bottom to the sea surface. The black 
dot denotes the substellar point. In these simulations, the stellar temperature is 3,400 K, the stellar 
flux is 1,400~W\,m$^{-2}$, and the planetary rotation period is 37 Earth days (see Table~1).}
\label{Figure:OceanDepthRidges}
\end{center}
\end{figure}

An important result of this study is that the location of the habitable zone’s inner edge 
should not depend significantly on ocean dynamics. This is consistent with \cite{Wayetal2018Ocean}, 
who found that the effect of ocean dynamics on the climate decreases rapidly with stellar flux for 
a variety of rotation rates in the ROCKE3D GCM (their Fig.~2). We should note, however, 
our experiments cover a limited range of 
planetary parameters. We have examined the dynamics of an aqua-planet ocean having a depth 
of $\simeq$\,4,000~m. Sensitivity tests show that at a stellar flux of 1,400~W\,m$^{-2}$ for 
shallower oceans or for oceans surrounded 
by continents, the day-to-night OHT would become smaller and its climatic effect would be weaker 
(Fig.~\ref{Figure:OceanDepthRidges}). For a shallower ocean, friction at the ocean bottom 
is more effective at decelerating ocean currents, 
so that less warm substellar seawater can be transported to the cold night side. 
In the case of an 800-m deep ocean, the night side is much cooler than that in the 4,000-m case 
as a result of less ocean heat input from the day side. The dayside surface, however, also 
becomes cooler even though less energy is transported away from the substellar region. 
This is due to a cloud feedback. Due to the reduced OHT (Fig.~\ref{Figure:OceanDepthRidges}(d)), 
the surface temperature contrast between the day side and the night side increases 
(Fig.~\ref{Figure:OceanDepthRidges}(c)). The larger surface temperature contrast promotes stronger 
water vapor convergence above the substellar region, and therefore more clouds form there. 
These clouds have a strong cooling effect on the sea surface through increasing planetary albedo. 
The planetary albedos are 0.29 and 0.45 in the cases of 4,000 and 800 m, respectively. 
When the ocean depth is 400 m, the zonal (West-East) 
ocean currents become weaker (compared to the 800 m case), so that less energy is transported  
to the night side through ocean dynamics (Fig.~\ref{Figure:OceanDepthRidges}(f));  
however, the meridional (North-South) oceanic heat transport 
increases, so that the high latitudes become warmer (Fig.~\ref{Figure:OceanDepthRidges}(e)), similar to the results of \cite{Delgenioetal2017:Proxima_b}. 
Compared the 4,000-m aqua-planet, the trends in the one barrier case and the two barriers 
case are similar to the 800-m aqua-planet case: The day-to-night ocean heat transport reduces 
and both dayside and nightside surfaces become cooler (Fig.~\ref{Figure:OceanDepthRidges}(g-h)). 
In the two barriers case, the day-to-night OHT is completely blocked by the continents and the ocean 
currents can transport the substellar heat only to the terminators and to the day-side polar regions, 
where the surface becomes warmer than in the one barrier case (Fig.~\ref{Figure:OceanDepthRidges}(i-j)). 
Future work is required to further investigate the ocean barrier cases, 
as well as more ocean depth cases, under higher stellar fluxes.

Future work is also required to investigate the effects of ocean dynamics on planets in 
different spin-orbit resonance states (such as 3:2 for Mercury), on rapidly rotating planets 
around G stars (such as Earth) and on planets near the outer edge of the habitable zone, 
as well as the effects of different atmospheric masses, atmospheric compositions and of 
deeper oceans \citep[such as ocean worlds,][]{Legeretal203}. A higher background atmospheric pressure 
than the one bar used in this study may further lessen the surface shortwave 
energy deposition through increasing atmospheric scattering, which may further 
weaken the effect of ocean dynamics on the inner edge of the habitable zone. A lower 
background atmospheric pressure should have a small effect on the results shown here 
because water vapor should dominate atmospheric composition at the inner edge of the 
habitable zone, especially for the runaway greenhouse state.

In our experiments, the salinity of sea ice is set to 4 g kg$^{-1}$ while the seawater 
salinity is about 35 g kg$^{-1}$, so that sea ice melting and freezing is able to 
influence the ocean salinity and thereby oceanic thermohaline circulation. 
The thermohaline circulation has not been shown in this manuscript because 
the day-to-night OHT is dominated by wind-driven ocean circulations 
in our experiments. Our model blew up when the stellar flux 
was higher than 2,800~W\,m$^{-2}$. Future experiments are required for higher stellar fluxes, 
under which wind-driven ocean circulation would become even weaker and thermohaline 
circulation might become important. Our preliminary thinking is that for planets near 
the inner edge of the habitable zone, the thermohaline circulation should be 
mainly driven by the salinity gradient because the surface temperature gradient is 
very small. The salinity gradient is mainly determined by the spatial pattern 
of precipitation and evaporation that are driven by large-scale atmospheric circulation. 
For planets at the outer edge of the habitable zone, the thermohaline circulation 
may be larger than that at the inner edge. 
This is because both temperature and salinity gradients on planets at the outer edge 
may be strong enough to drive a robust thermohaline circulation.
\cite{Cullumetal2014rotation,Cullumetal2016salinity} examined 
the effects of planetary rotation and average ocean salinity on the thermohaline 
circulation using an ocean-only model. Wind-driven ocean circulation and the 
interactions between ocean, atmosphere and surface climate, however, were not 
considered in their studies. Future work is required to examine the effect of 
average ocean salinity on the ocean circulation (as was briefly tested 
in \cite{Delgenioetal2017:Proxima_b} for Proxima b) and the effect of thermal-driven 
ocean circulation on the inner and outer edges of the habitable zone.


\section{Summary}
\label{sec:Summary}

\begin{figure}[!htbp]
\begin{center}
\includegraphics[angle=0, width=16cm]{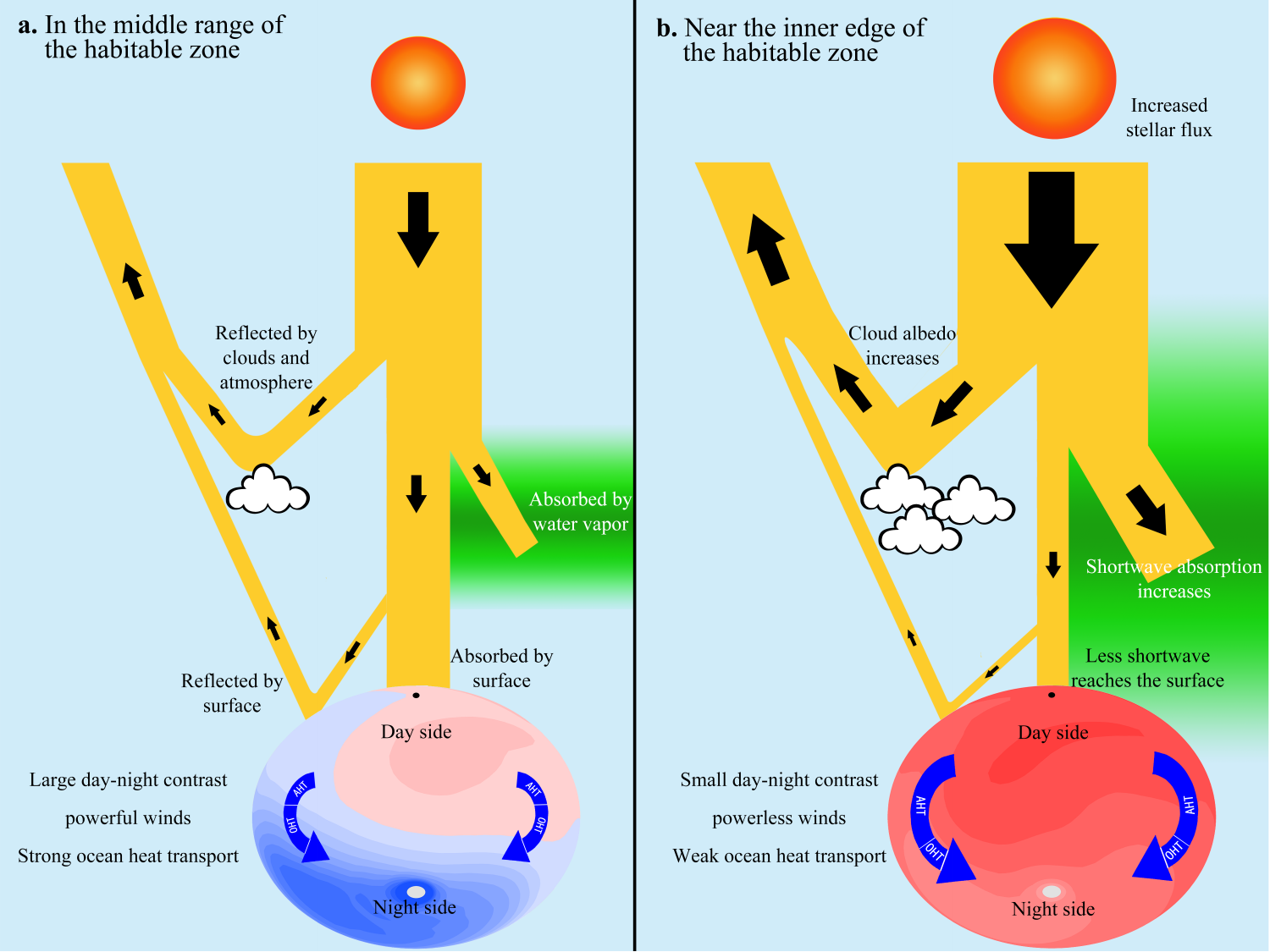}
\caption{Schematic illustration of the decrease in day-to-night ocean heat transport (OHT) as 
the stellar flux increases and the underlying mechanisms. The decreasing trend of OHT results from 
weaker wind stress combined with less stellar radiation depositing energy at the dayside surface. 
(a): In the middle range of the habitable zone, the day-night surface temperature contrast is 
strong, the surface wind stresses are robust, stellar radiation absorbed by water vapor or reflected 
by clouds is relatively small, a large part of the stellar energy reaches the surface, and the ocean 
dominates the day-to-night heat transport. (b): Near the inner edge of the habitable zone, the surface is 
hot, the day-night surface temperature contrast is small, the surface wind stresses are weak, stellar 
radiation absorbed by water vapor or reflected by clouds is relatively large, a small part of the stellar 
energy reaches the surface, and the atmosphere dominates the day-to-night heat transport.}
\label{Figure:Summary}
\end{center}
\end{figure}

In the middle range of the habitable zone, 
ocean dynamics significantly warm the night side and the dayside high latitudes of 
1:1 tidally locked aqua-planets and produce an eastward shift of the hottest point at the surface. 
For planets near the inner edge of the habitable zone, however, oceanic heat transport 
is weak and has nearly no effect on the location of the inner edge or on the thermal phase 
curves of planets near the inner edge (summarized in Fig.~\ref{Figure:Summary}). 
The weakening of oceanic heat transport with increasing stellar flux is due to the 
combined effect of weakened surface wind stress and decreased surface stellar 
energy deposition at the sea surface. Atmospheric heat transport increases with stellar flux and 
dominates on planets near the inner edge of the habitable zone. Finally, we note 
that the detection of oceans, continents and atmospheres on distant terrestrial exoplanets 
is still a big challenge \citep[e.g.,][]{Cowan:2009p3255,cowan12-glint,cowan12-thermal}. 
Future observations using high temporal frequency specular reflections as well as emission and 
transmission spectra may be able to infer the surface as well as atmospheric characteristics 
of nearby transiting planets \citep{Cowanetal2015,Greeneetal2016,Stevensonetal2016,Beanetal2018}.


\acknowledgments \textbf{Acknowledgments:} 
We are grateful to Junyan Xiong for his help in drawing Figure~\ref{Figure:Summary}. 
J.Y. acknowledges support from the National Science Foundation 
of China (NSFC) grants 41861124002, 41675071, 41606060, and 41761144072.




\bibliography{ocean_inner_edge_20190206.bbl}

\end{document}